\documentclass{aastex}

\shorttitle{Variability of S5 0716+714}
\shortauthors{Dai et al.}

\begin{document}
\title{Seven-Year Multi-Color Optical Monitoring of \\ BL Lacertae Object S5 0716+714}

\author{Yan Dai}
\affil{Department of Astronomy, Beijing Normal University, \\ Beijing 100875, China \\
Department for Popularization of Astronomy, Beijing Planetarium, \\
138 Xizhimenwai Street, Beijing 100044, China}

\author{Jianghua Wu, Zong-Hong Zhu}
\affil{Department of Astronomy, Beijing Normal University, \\ Beijing 100875, China}
\email{zhuzh@bnu.edu.cn}

\author{Xu Zhou, Jun Ma}
\affil{Key Laboratory of Optical Astronomy, National Astronomical Observatories, \\
Chinese Academy of Sciences, 20A Datun Road, Beijing 100012, China}

\author{Qirong Yuan}
\affil{Department of Physics and Institute of Theoretical Physics, Nanjing Normal University, \\ Nanjing 210046, China}

\and

\author{Lingzhi Wang}
\affil{Department of Astronomy, Beijing Normal University, \\
Beijing 100875, China}

\begin{abstract}

We have monitored the BL Lac object S5 0716+714 in five intermediate
optical wavebands from 2004 September to 2011 April. Here we present
the data that include 8661 measurements. It represents one of the
largest databases obtained for an object at optical domain. A simple
analysis of the data indicates that the object was active in most time,
and intraday variability was frequently observed. In total, the object
varied by 2.614 magnitudes in the $i$ band. Strong bluer-when-brighter
chromatism was observed on long, intermediate, and short timescales.

\end{abstract}


\section{Introduction}

Blazars constitute the most variable subclass of active galactic
nuclei (AGNs). Depending on whether or not showing strong emission
lines in spectra. Blazar is divided into flat-spectrum radio
quasars (FSRQs) and BL Lacertae (BL Lac) objects. BL Lac objects are
characterized by non-thermal continuum emission across the whole
electromagnetic spectrum with absent or weak emission and absorption
lines \citep{Sti93}, variable and high polarization
\citep{Ang80,Imp88,Gab89}, large amplitude and rapid variability at
all wavelengths from radio to gamma rays \citep{Rav02,Bot03}, and
superluminal motion of radio components \citep{Den00}.

S5 0716+714 is a distant BL Lac. In 1979, it was discovered in a survey
for sources with a 5 GHz flux greater than 1 Jy \citep{Kuh81}. Because
of the featureless spectrum and its strong optical polarization, it was
identified as a BL Lacertae object by \citet{Bie81}. The redshift
of S5 0716+714 was uncertain until Wagner et al. (1996) estimated
a value bigger than 0.3. Afterwards, Nilsson et al. (2008) acquired
a deep i-band image of this object and derived a redshift of
$0.31\pm 0.08$ by using the host galaxy as a ¡¯standard candle¡¯.
Most recently, Danforth et al. (2012) set an upper bound of $z < 0.304$
with a confidence level of 90\% for this object. This source is one
of the most studied BL Lac objects, because it has high brightness
and strong variability. The optical duty cycle of S5 0716+714 is
nearly unity, indicating that the source is always in an active
state in the visible \citep{Wag95}. Strong bluer-when-brighter
correlations were found for both internight and intranight variations
\citep{Wu05,Wu07,Wu12,Poo09,Hao10,Cha11}.

This source has been intensively monitored by a number of authors.
During a 4-week period of continuous monitoring, the source
displayed in both optical and radio regimes a transition between
states of fast and slow variability with a change of the typical
variability timescale from about 1 to about 7 days \citep{Qui91}.
Wagner et al. (1996) investigated the rapid variations of this
object in the radio, optical, ultraviolet, and X-ray regimes and
found that it always keeps high amplitude change on the timescale
of a few days. Sagar et al. (1999) showed an average $V-R$ color
of this BL Lac to be $\sim$0.4 mag in their one month long $BVRI$
optical monitoring campaign in 1994. \citet{Rai03} reported that
the long-term optical brightness variations of this source appear
to have a characteristic timescale of 3.3yr and four major optical
outbursts were observed at the beginning of 1995, in late 1997, at
the end of 2000, and in fall 2001. In particular, an exceptional
brightening of 2.3 mag in 9 days was detected in the $R$ band on
2000 October 30. Color analysis on the optical light curves reveals
only a weak general correlation between the color index and the source
brightness. Recently, Poon et al. (2009) monitored the BL Lac object
S5 0716+714 in the optical band during 2008 October and December and
2009 February with a best temporal resolution of about 5 minutes in
the $BVRI$ bands. Typical timescales of microvariability range from
2 to 8 hr. The overall $V-R$ color index ranges from 0.37 to 0.59.
Strong bluer-when-brighter chromatism was found on internight
timescales. The overall variability amplitude decreases with decreasing
frequency.

We have monitored S5 0716+714 since 2004. Here we present the data
during the period from 2004 to 2011. A simple analysis is performed
and the results are described.

This paper is organized as follows. The Observation and data
analysis is described in Section 2. Section 3 presents the light curves.
Section 4 shows the comparison result of $i$-data from us and $R$-data
from other authors. The relation of color and magnitude is described in
Section 5. The conclusions are given in Section 6.

\section{Observations and data analysis}

Our optical monitoring program of S5 0716+714 was carried out with
the 60/90 cm Schmidt telescope located at the Xinglong Station of
the National Astronomical Observatories of China (NAOC). Prior to
2006, a Ford Aerospace 2048$\times$2048 CCD camera was mounted at
its main focus. The CCD has a pixel size of 15 $\mu$m, and its field
of view is 58'$\times $58', resulting in a resolution of 1\farcs7
pixel$^{-1}$. At the beginning of 2006, the 2k CCD was replaced by
a new 4096$\times$4096 CCD. The field of view is now 96'$\times$96',
resulting in a resolution of 1\farcs3 pixel$^{-1}$. The telescope
is equipped with 15 intermediate-band filters, covering a wavelength
range from 3000 to 10000 {\AA}.

This paper includes data from 2004 September 10 to 2011 April 24.
Excluding the nights with bad weather and those devoted to other
targets, the actual number of nights for S5 0716+714 observations
is 332. We used filters in $e$, $i$, and $m$ bands to observe in
2004-2006, and then changed to the $c$, $i$, and $o$ bands from
2006 December. The central wavelengths of the $c$, $e$, $i$, $m$,
and $o$ bands were 4210, 4920, 6660, 8020, and 9190 {\AA}, respectively.
The central wavelength of $i$ band is similar to the $R$ band.
With the observational results of stars, the magnitudes in these
two bands can be transformed with $R=i+0.1$ \citep{Zho03}.
Depending on the weather and seeing conditions, the exposure time
of different bands range from 30s to 480s, and the exposures per
night of different bands varies between 2 to 43.

The data reduction procedure includes bias subtraction,
flat-fielding, extraction of instrumental aperture magnitude, and
flux calibration. We used differential photometry. For each frame,
the instrumental magnitudes of the blazar and four comparison stars
(See Fig.1) were extracted at first. The radii of the aperture and
the sky annuli were adopted as 3, 7, and 10 pixels, respectively.
Then the brightness of the blazar was measured relative to the
average brightness of the three reference stars 3, 4, and 5. Star 6
acted as a check star, which has an apparently similar brightness as
the blazar \citep[for a reasonable selection of reference and check
stars, see][]{How88}. The differential magnitude of star 6 is the
difference between the magnitude of star 6 and the average magnitude
of star 4 and 5, so as to verify the stable fluxes of the four comparison
stars, and to verify the accuracy of our measurements. The $c$, $e$,
$i$, $m$, and $o$ magnitudes of the 4 comparison stars were obtained
by observing them and the standard star HD 19945 on a photometric night
and are listed in Table 1.

\section{Light Curves} \label{bozomath}

The samples of observational log and results are given in Tables
2-6. The columns are observation date and time in universal time,
Julian date, exposure time in second, magnitude and error of S5
0716+714, and differential magnitude of star 6 (its nightly
averages were set to zero). Figure 2 shows the light curves of the
overall monitoring period in the five bands.

The source remained active during the whole monitoring period. The
variation amplitudes of $e$, $i$, and $m$ bands from 2004 September
10 to 2006 March 29 are 1.200, 1.156, and 1.127 mags, respectively,
and the variation amplitudes of $c$, $i$, and $o$ bands from 2006
December 6 to 2011 April 24 are 2.763, 2.614, and 2.522 mags,
respectively. The amplitude of variation tends to decrease with
decreasing frequency. The light curves keep fluctuating during the
whole monitoring period. The curves get the extreme bright value
on 2004 September 10 (JD 2,453,259), 2007 October 20 (JD 2,454,394),
2008 April 22 (JD 2,454,579), and 2010 September 28 (JD 2,455,468),
and get the extreme dark value on 2005 January 29 (JD 2,453,400),
2007 December 16 (JD 2,454,451), and 2011 April 24 (JD 2,455,676).
A faintest optical state was recorded on 2007 December 16 (JD 2,454,451),
. Nilsson et al. (2008) acquired a deep i-band image of this object
at that state and derived a redshift of $0.31\pm 0.08$ by using the
host galaxy as a 'standard candle'.

The BL Lac object S5 0716+714 is one of the brightest BL Lac objects
noted for its microvariability. In order to confirm whether or not
the object was variable in one day, a quantitative assessment was
carried out. For each of the 233 nights with observational duration
longer than 2 hours, we used a chi-square inspection to check whether
there is intraday variability (IDV) \citep{Pen70,Kes76,de10}. The
chi-square value was compared with the critical value at the 95\%
confidence level. If the former was greater than the latter, the null
hypothesis that there was no variability was rejected. As a result,
138 nights (62\% of 233) with IDV were identified. The observed fastest
variation of S5 0716+714 varied by 0.117 mags in 1.1 hrs in the $c$ band
on 2011 March 5 (JD 2,455,626), as indicated by the bottom line with
arrows in Figure 3. The similar work was made by Villata et al. (2000).
They found that the source exhibited strong variability with similar
trend but different amplitudes in all bands, and noted the monotonic
brightness increase in $B$ band for about 130 minutes. The steepest
(linear) part has a rising rate of 0.002 mag per minute and a duration
of about 45 minutes.

\section{Compare with Other Data} \label{bozomath}

In Section 2, a transforming formula between the $i$ and $R$ magnitudes
was mentioned. This formula was derived from the observational result of
stars (Zhou et al. 2003). However, the spectral shape of blazars is quite
different from that of stars. So the transforming formula may be different
for blazars. Therefore, we made a comparison between our $i$-band data and
the $R$-band data of other authors in order to find an empirical relation
between them. Villata et al. (2000) and Poon et al. (2009) have made intensive
monitoring on the same object. Their data were adopted and matched in time
with ours with a threshold of less than 0.02 days (or 28.8 minutes). As the
results, 24 $R$-$i$ matches were found for Villata et al. and our data, and
97 $R$-$i$ matches were found for Poon et al. and our data. The average time
differences are 0.0038 and 0.0009 days for the 20 and 97 matches, respectively.
Two $R$-$i$ diagrams were plotted in the top and middle panels of Figure 4.
Two linear regressions give the transforming formulae as
$R=(0.964\pm0.015) \times i + (0.279\pm0.187)$ and $R=(0.976\pm0.009) \times i + (0.059\pm0.123)$,
respectively. If all matches are plotted together, as shown in the bottom
panel of Figure 4, the linear regression gives the formula as
$R=(0.897\pm0.004) \times i + (1.127\pm0.048)$.

\section{Relation of Color and Magnitude} \label{bozomath}

The long-term color behavior of S5 0716+714 was studied based on our
data. For the 2004-2006 data, the $e-m$ color was calculated and plotted
vs. the $e$ magnitude in the top panel of Figure 5. For the data after
2006 December, the $c-o$ color was calculated and plotted vs. the $c$
magnitude in the central panel of Figure 5. The bottom panels illustrate
how the $c-o$ color and $c$ magnitude changed with time. Despite the
discontinuity in the top panel and significant scatter in the central
panel, there is an overall bluer-when-brighter chromatism in both panels.

In order to investigate the color behavior of S5 0716+714 on the intermediate
timescale, three episodes in our monitoring are isolated. They are from
JDs 2,454,101 to 2,454,115, from JDs 2,454,429 to 2,454,463, and from
JDs 2,455,597 to 2,455,629. These three episodes lasted from two weeks
to more than one month. In these periods, we have relatively continuous
monitoring, and the object showed significant variations. The light
curves and the corresponding color-magnitude diagrams are plotted in
Figure 6 for the three episodes. There are strong color-magnitude
correlations. The correlation coefficients are 0.89, 0.78, and 0.82,
respectively. The bluer-when-brighter chromatism on intermediate
timescale is found by other authors \citep[e.g.,][]{Vil00,Wu07}. On intraday
timescales, the object also displays strong bluer-when-brighter
chromatism. Some examples are given in Figure 7.

Depending on the balance between escape, acceleration, and cooling of
the electrons with different energy, either soft (low energy) or hard
(high energy) lags are expected \citep{Kir98}. This will lead to a
loop-like path of the blazar¡¯s state in a color-magnitude (or spectral
index-flux) diagram. The direction of this spectral hysteresis can be
either clockwise or anticlockwise. It depends on the relative position
between the observing frequency and the peak frequency of the synchrotron
component in the SED of the blazar as well as the relative values of the
acceleration, cooling and escape timescales \citep{Chi99,Der98}.

In our monitoring, we got an inconspicuous loop flare in internight
timescale and an obvious loop flare in intranight timescale, which
are shown in Figure 8. The upper-left panel is the light curve of JDs
2,454,390 $\sim$ 2,454,398, which shows a flare in $cio$ bands. The
corresponding color-magnitude diagram is displayed in the upper-right
panel, in which the numbers denote time sequence. There were not
significant variations during these day, so we averaged the data in
day to decrease system error. The points spread as diagonal distribution
in color-magnitude diagram. The loop flare is far less obvious. The result
of intranight timescale, JD 2,455,621, is shown in the bottom of Figure 7.
The numbers denote the time sequence. An anticlockwise loop can be seen
in the lower-right panel. As Kirk et al. (1998) supposed, if the loop is
traced anticlockwise, there might be a flare propagating from lower to
higher energy, as particles are gradually accelerated into the radiating
window. The frequency of $c$ and $o$ band are $7.13\times10^{14}\,$Hz
and $3.26\times10^{14}\,$Hz. In 1999, Giommi et al. found that the frequency
is between $10^{14}\,$ and $10^{15}\,$Hz. Our anticlockwise loop imply
that the peak frequency of short band in the SED of the blazar should
be higher than $7.13\times10^{14}\,$Hz.

\section{Conclusions}

We have monitored the BL Lac object S5 0716+714 in five intermediate
optical wavebands from 2004 September to 2011 April by the 60/90 cm
Schmidt telescope located at the Xinglong Station of the National
Astronomical Observatories of China (NAOC). We collected 8661 data
points with error less than 0.05 mags. It represents one of the largest
databases obtained for an object at optical domain and can be used to
study both the long- and short-term flux and spectral variability of
this object. It can also be correlated with the data in the radio, X-ray,
or gamma-ray wavelengths in order to investigate the broad-band behavior
of this object. A simple analysis of the data indicates that the object
was active in most time. The overall amplitudes of $e$, $i$, and $m$ bands
from 2004 September 10 to 2006 March 29 are 1.200, 1.156, and 1.127 mags,
respectively, and the overall amplitudes of $c$, $i$, and $o$ bands from
2006 December 6 to 2011 April 24 are 2.763, 2.592, and 2.522 mags,
respectively. The amplitude of variation tends to decrease with decreasing
frequency. The bluer-when-brighter phenomenon is effectively confirmed on
long, intermediate, and short timescales. It is an important support to
the shock-in-jet model in which shocks propagate down the relativistic
jet, accelerating particles and/or compressing magnetic fields, leading
to the observed flux and spectral variability \citep{Mar85,Qia91}. There
were 138 nights of IDV captured during the whole monitoring period.

\begin{acknowledgements}
The authors thank the anonymous referee for constructive suggestions
and insightful comments. We thank Villata, M. for kindly sending us
the WEBT data on S5 0716+714. This work has been supported by Chinese
National Natural Science Foundation grants 11273006, 11173016, and
11073023. WJH is supported by National Basic Research Program of China
£¨973 Program 2013CB834900£©. ZHZ acknowledges that this work is
supported by the Ministry of Science and Technology National Basic
Science Program (Project 973) under Grant No.2012CB821804, the
Fundamental Research Funds for the Central Universities and
Scientific Research Foundation of Beijing Normal University.
\end{acknowledgements}

\begin{figure}
\begin{center}
\includegraphics[width=14cm,height=14cm]{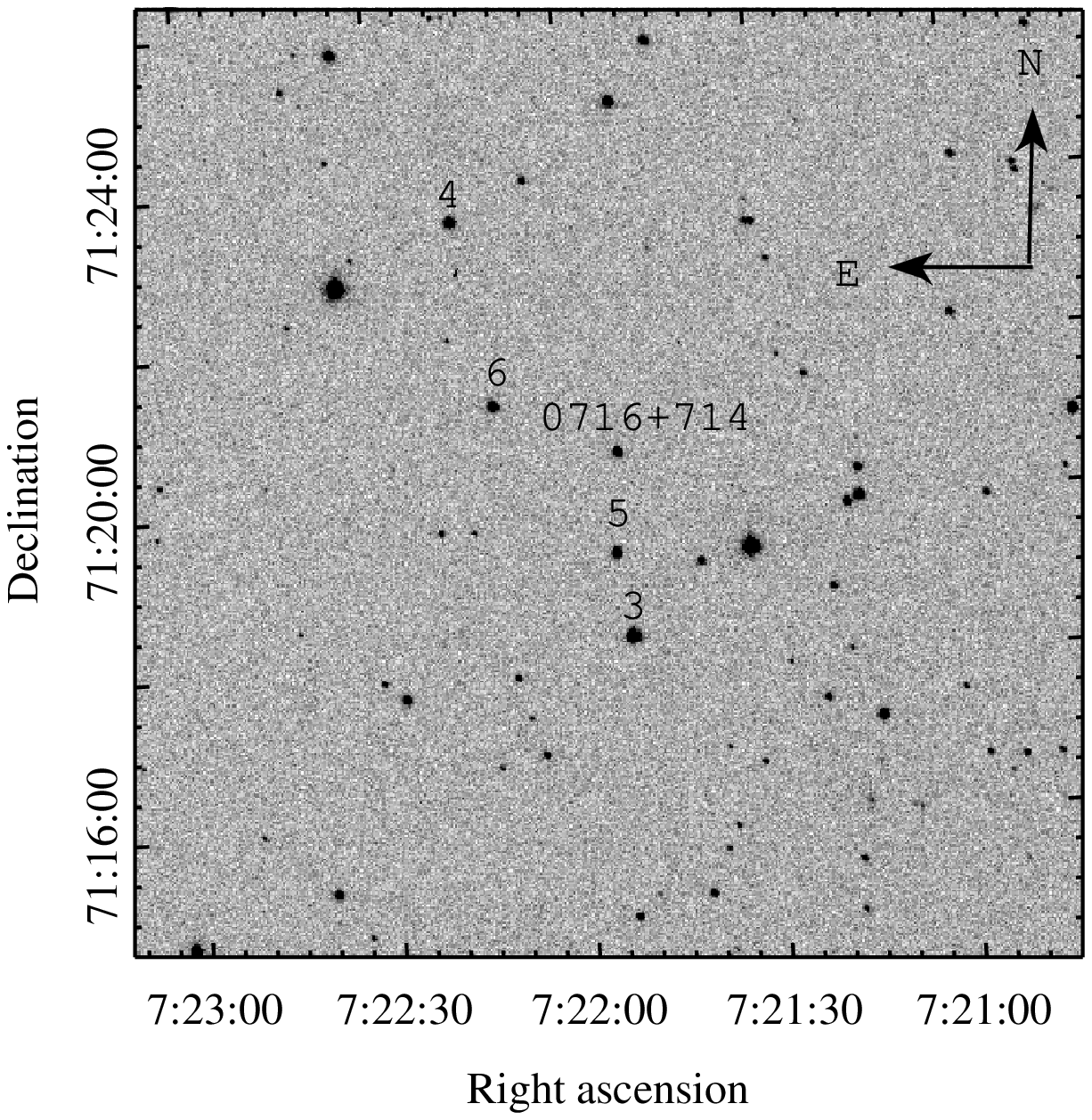}
\caption{Finding chart of S5 0716+714 and the 4 comparison stars taken
with the 60/90 Schmidt telescope and filter $i$ on 2011 April 12 (JD 2,455,664). The
size is $12'\times12'$ (or 512$\times$512 in pixels). \label{fig1}}
\end{center}
\end{figure}

\begin{figure}
\includegraphics[angle=0,width=1\textwidth]{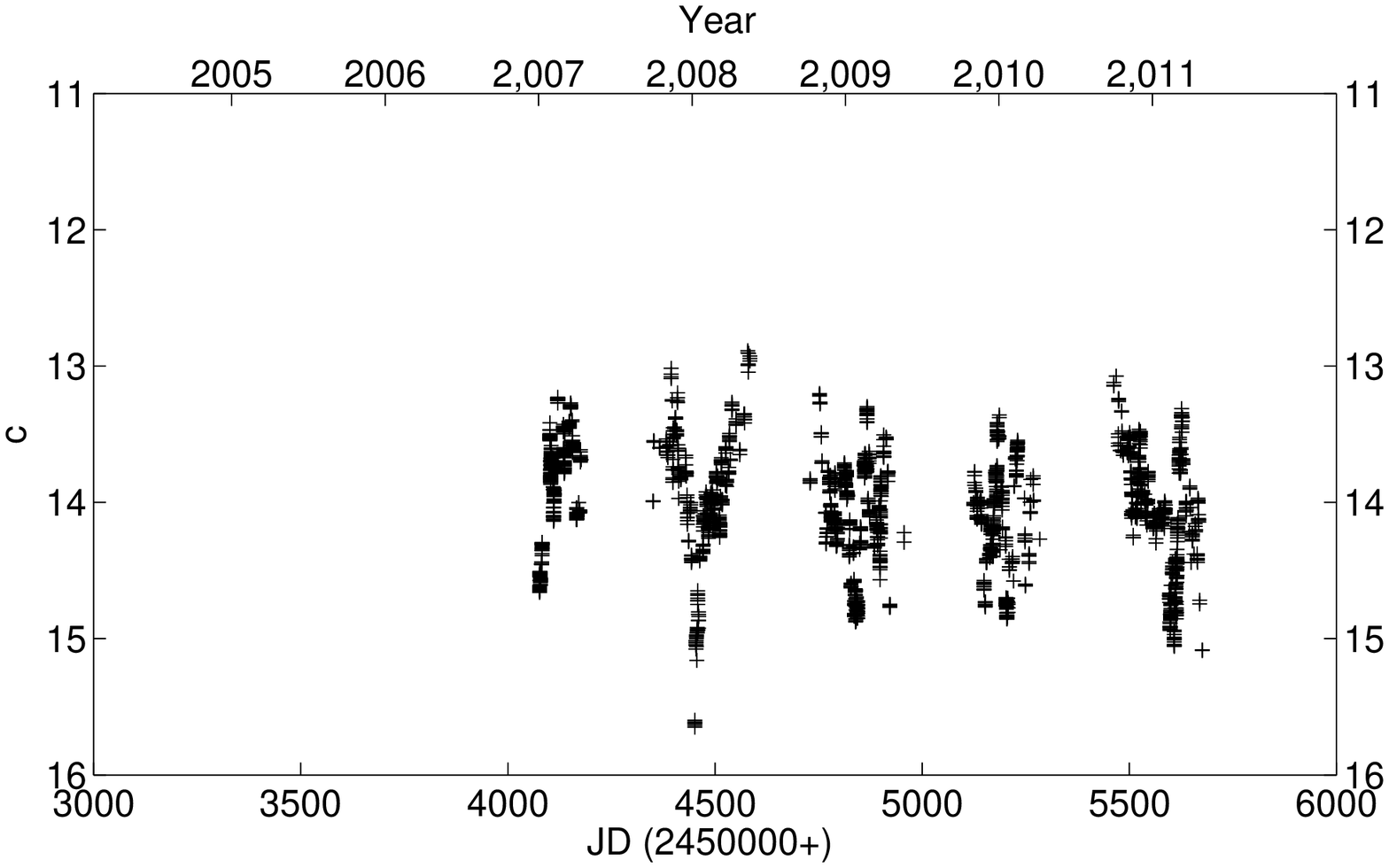}
\end{figure}
\begin{figure}
\includegraphics[angle=0,width=1\textwidth]{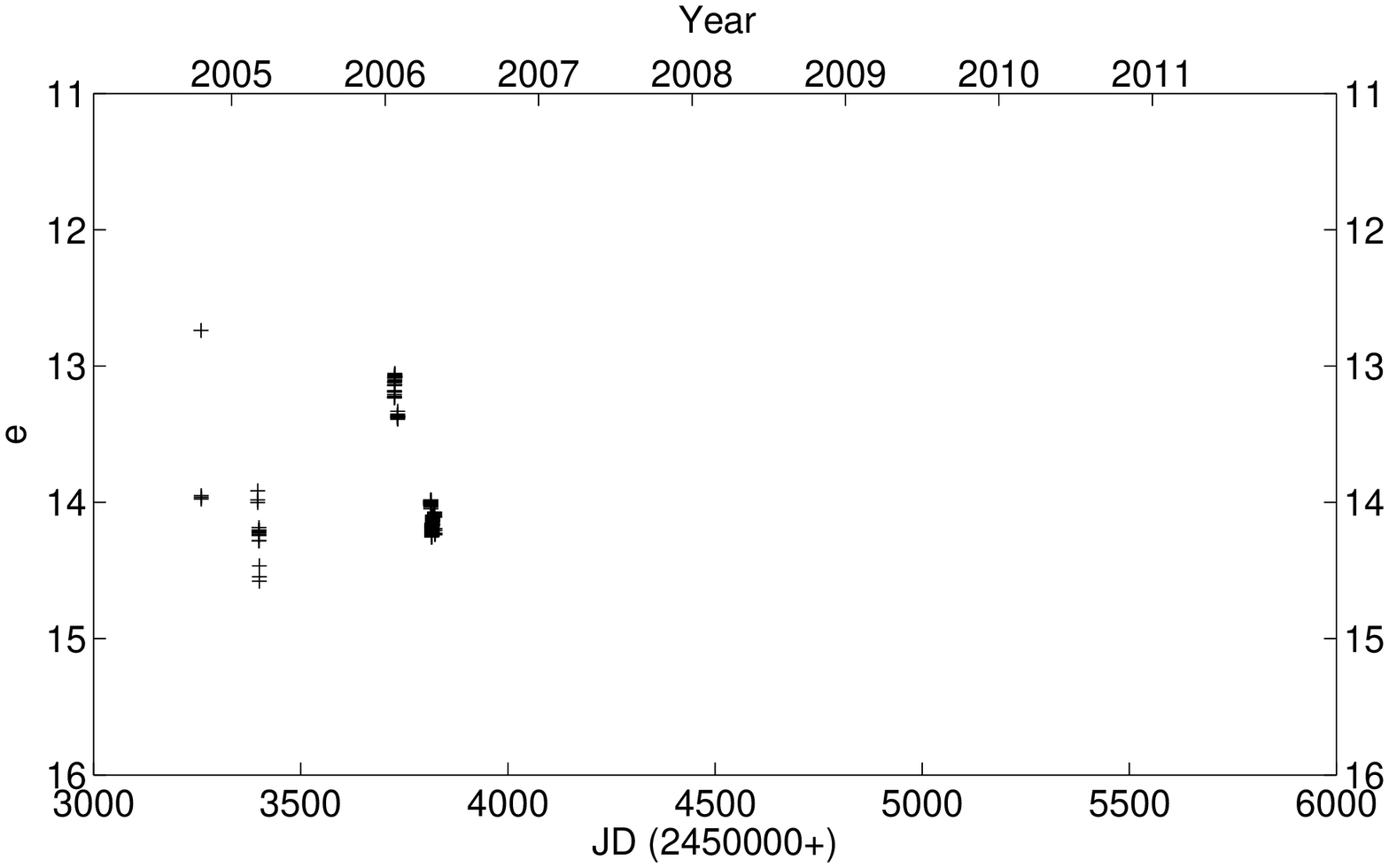}
\end{figure}
\begin{figure}
\includegraphics[angle=0,width=1\textwidth]{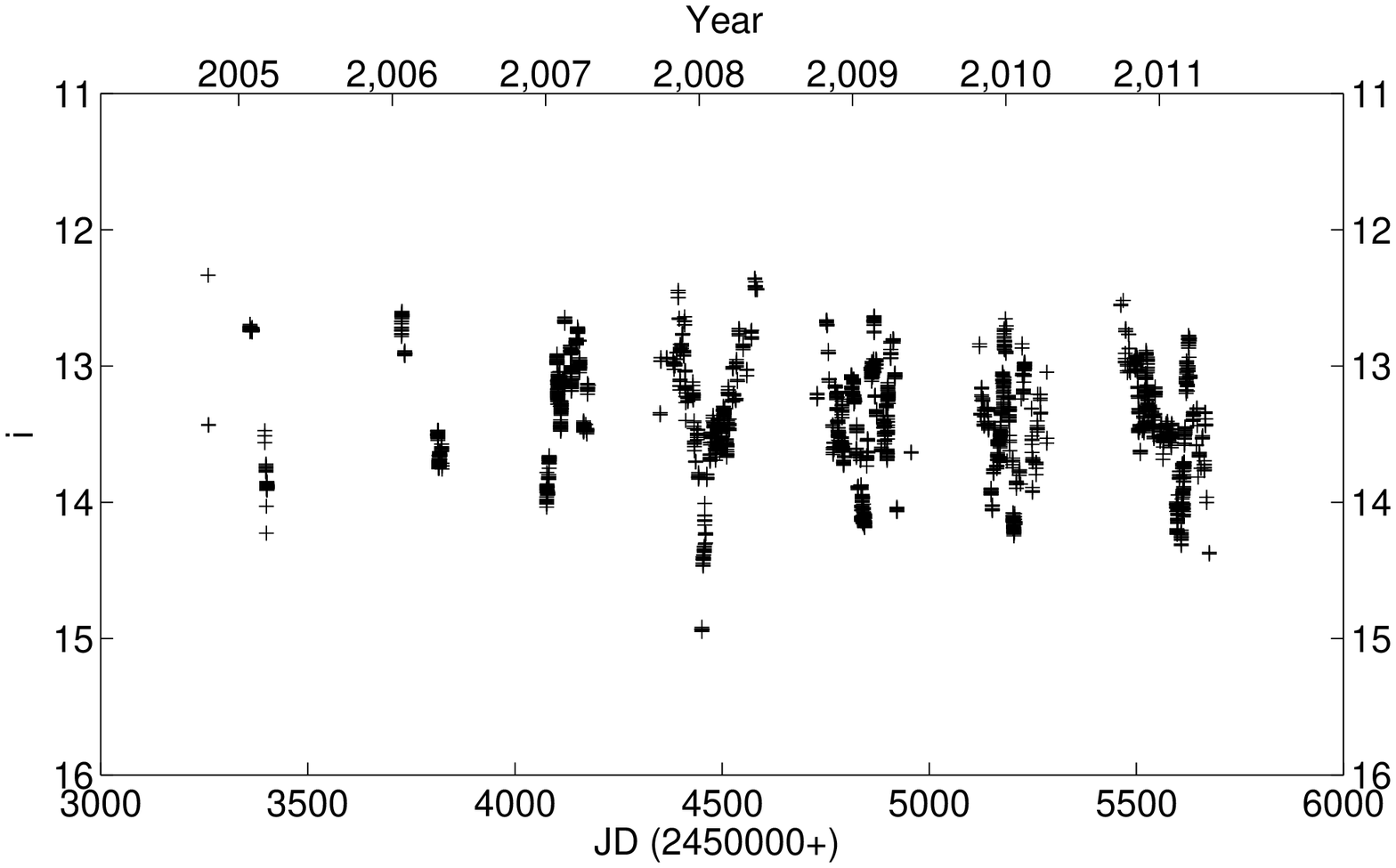}
\end{figure}
\begin{figure}
\includegraphics[angle=0,width=1\textwidth]{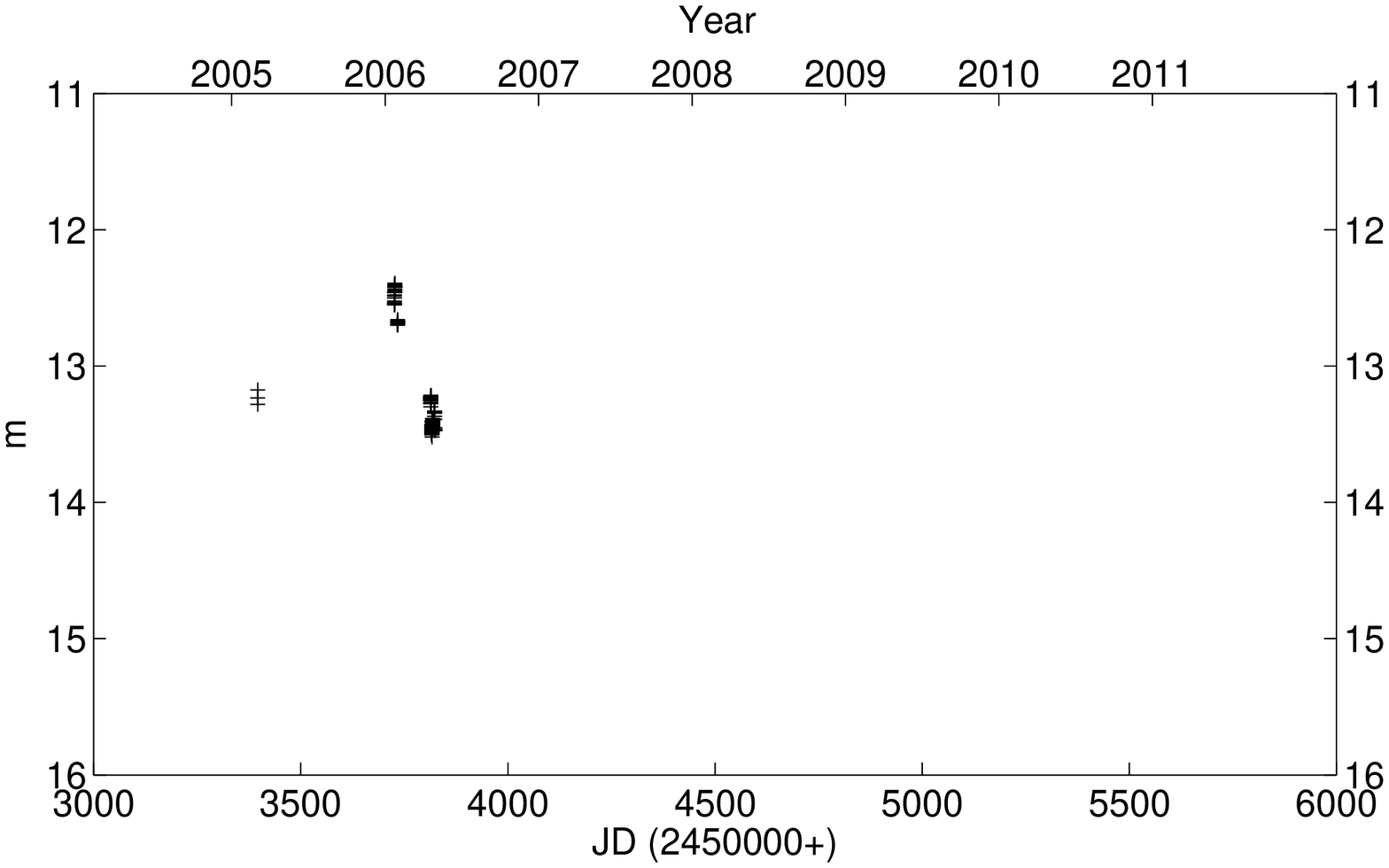}
\end{figure}
\begin{figure}
\includegraphics[angle=0,width=1\textwidth]{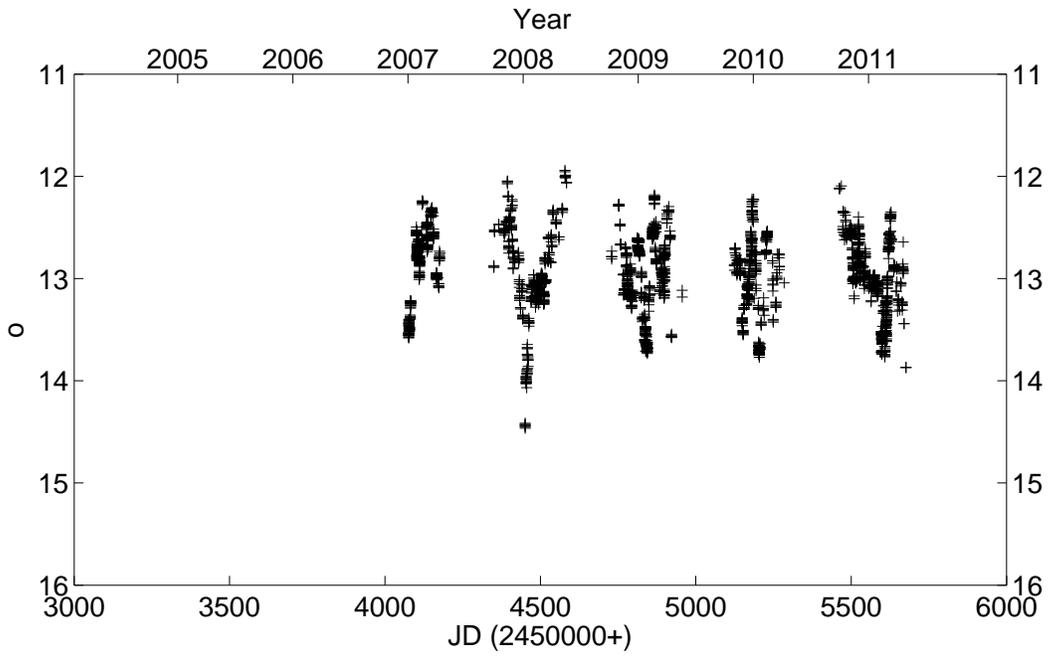}
\caption{Light curves of S5 0716+714 in the $c$, $e$, $i$, $m$,
and $o$ bands. \label{fig2}}
\end{figure}

\begin{figure}
\begin{center}
\includegraphics[width=14cm,height=14cm]{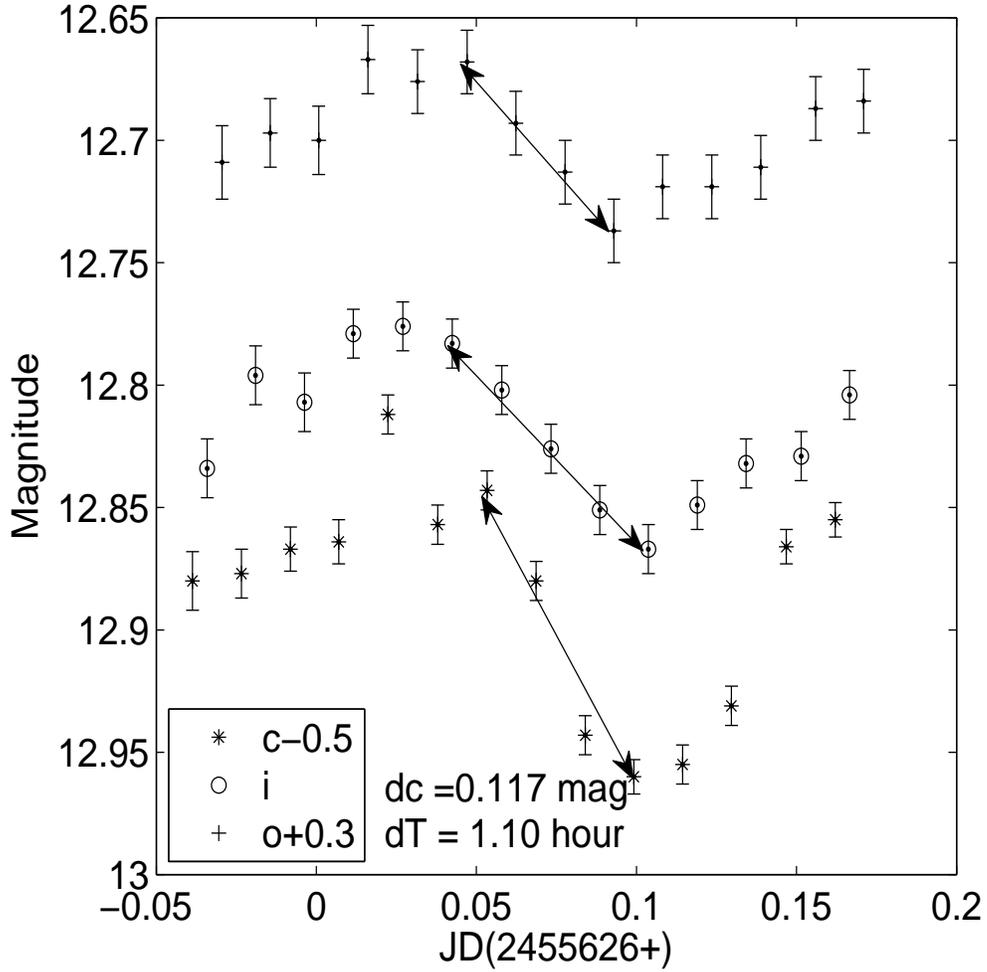}
\caption{The observed fastest variation of S5 0716+714. The
object varied by 0.117 mags in 1.1 hrs in the $c$ band, as
indicated by the bottom line with arrows. \label{fig3}}
\end{center}
\end{figure}

\begin{figure}
\begin{center}
\includegraphics[width=0.5\textwidth]{figure4_1_.ps}
\includegraphics[width=0.5\textwidth]{figure4_2_.ps}
\includegraphics[width=0.5\textwidth]{figure4_3_.ps}
\caption{$i-R$ diagrams between our $i$-band data and the $R$-band
data of Villata et al. (top), Poon et al. (middle), and both of them
(bottom). \label{fig4}}
\end{center}
\end{figure}

\begin{figure}
\begin{center}
\includegraphics[width=0.50\textwidth]{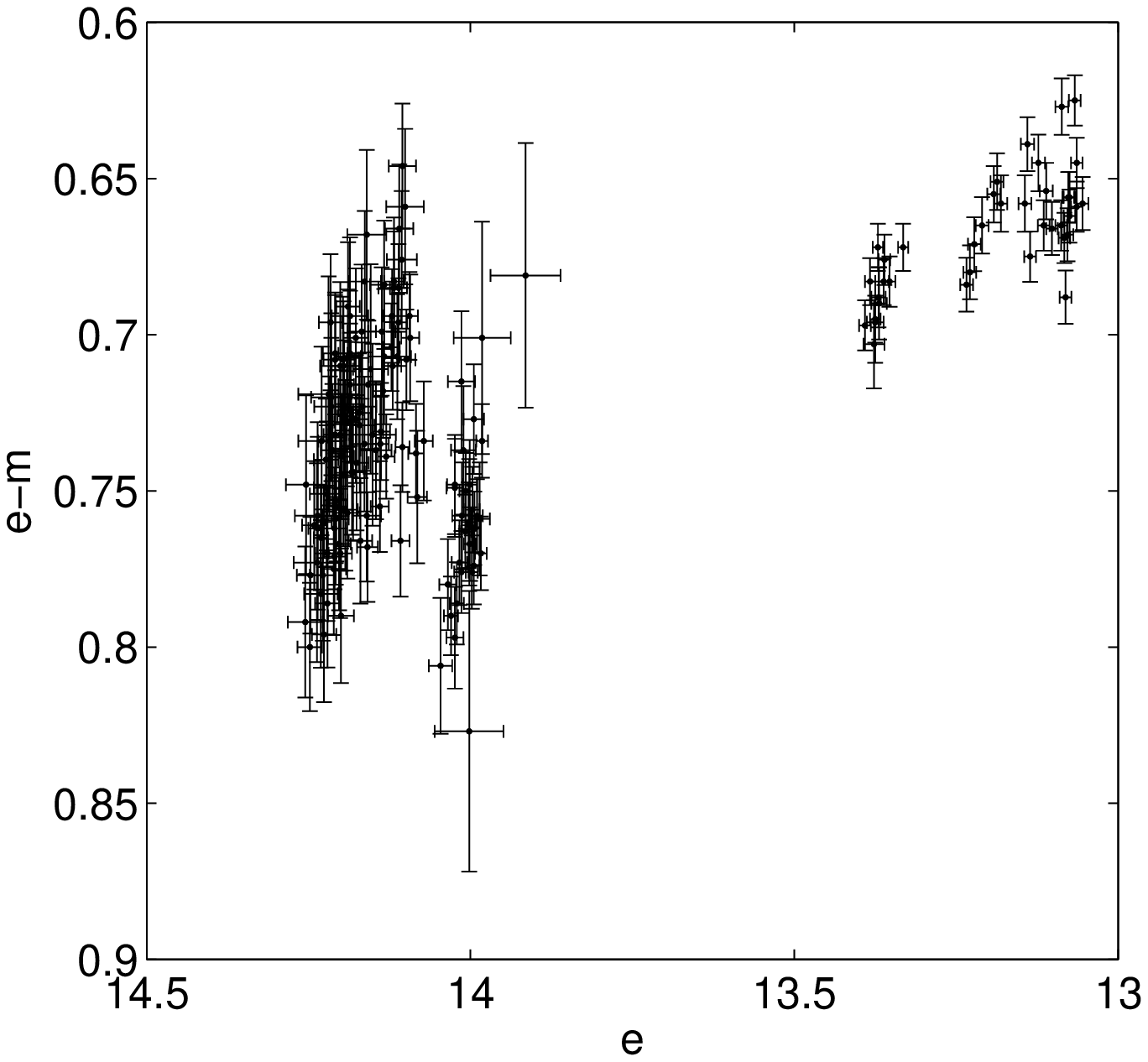}
\includegraphics[width=0.50\textwidth]{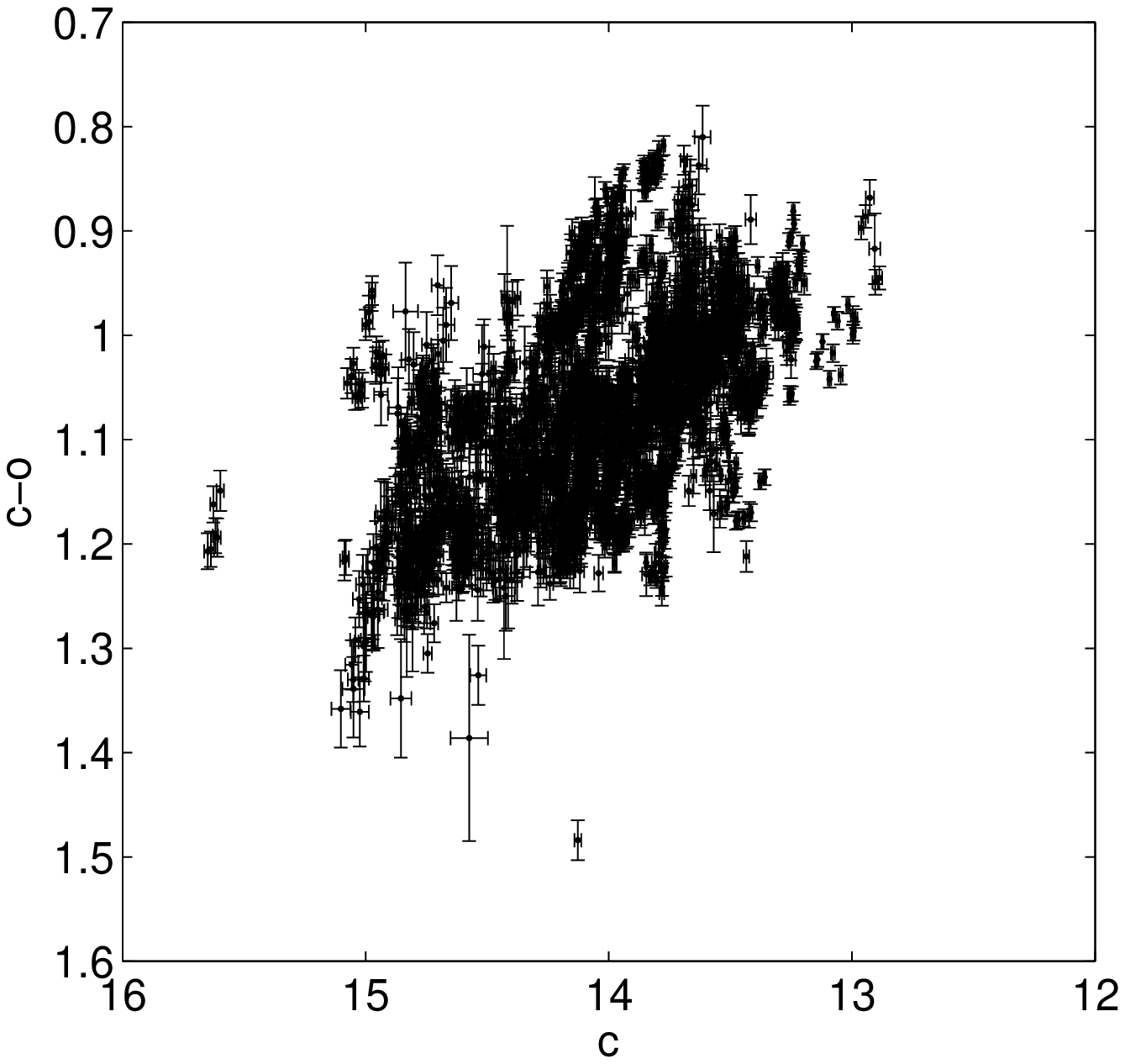}
\includegraphics[width=0.50\textwidth]{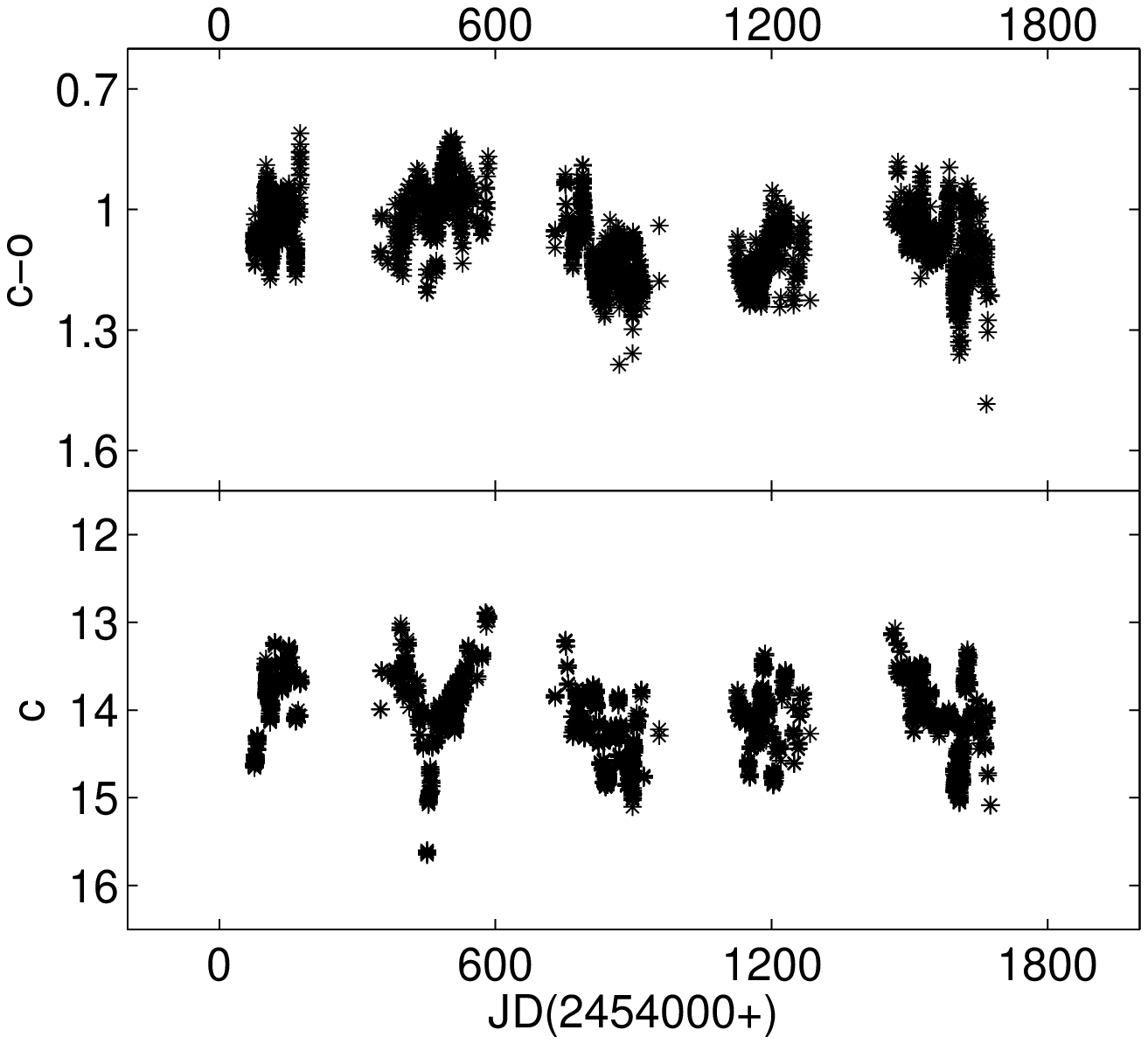}
\caption{Color-magnitude diagrams of $em$ bands (top) and $co$ bands
 (middle) and the color index and light curves of $co$ bands (bottom)
 during the whole monitoring period. \label{fig5}}
\end{center}
\end{figure}

\begin{figure}
\begin{center}
\includegraphics[angle=0,width=0.48\textwidth]{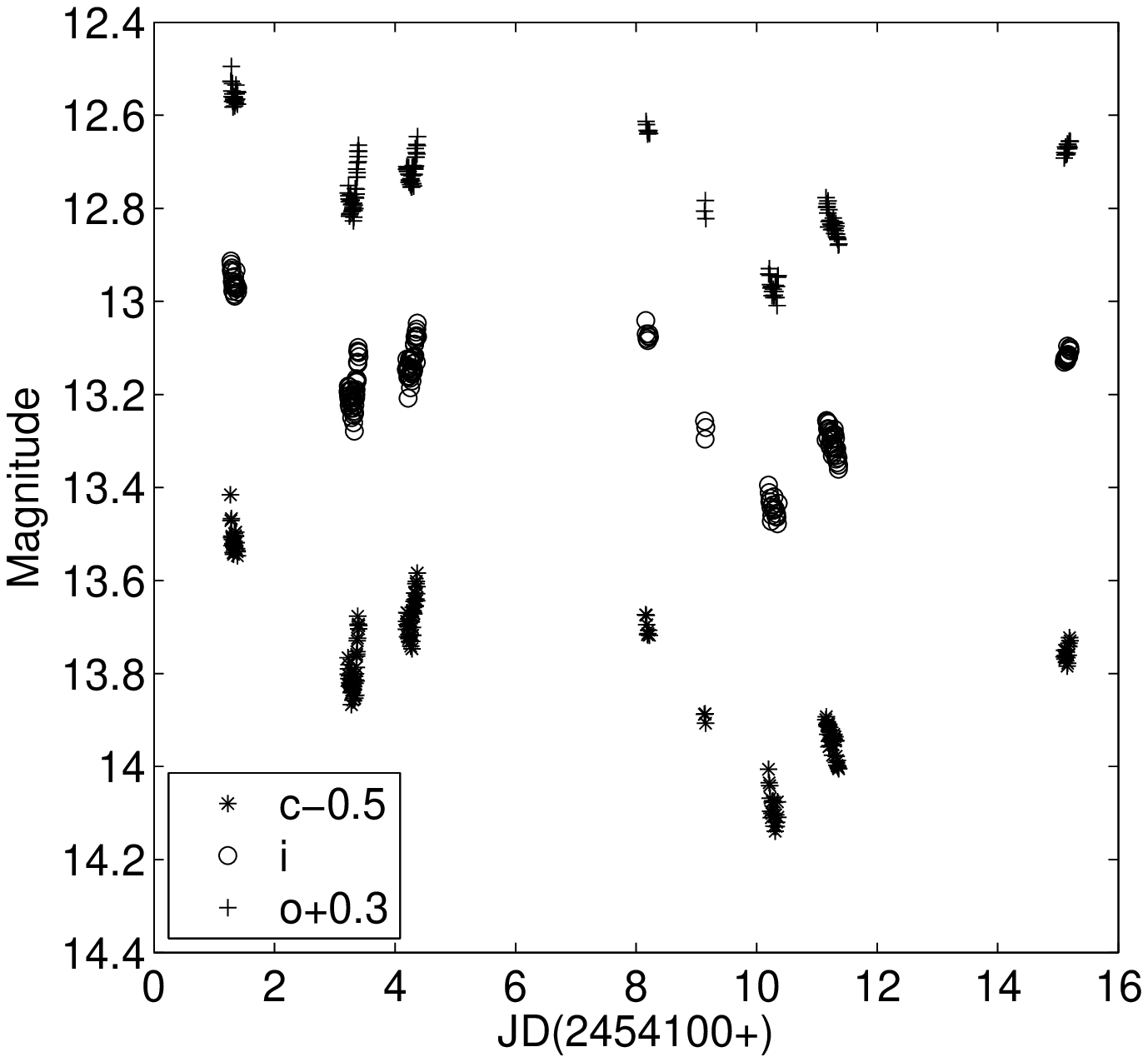}
\includegraphics[angle=0,width=0.48\textwidth]{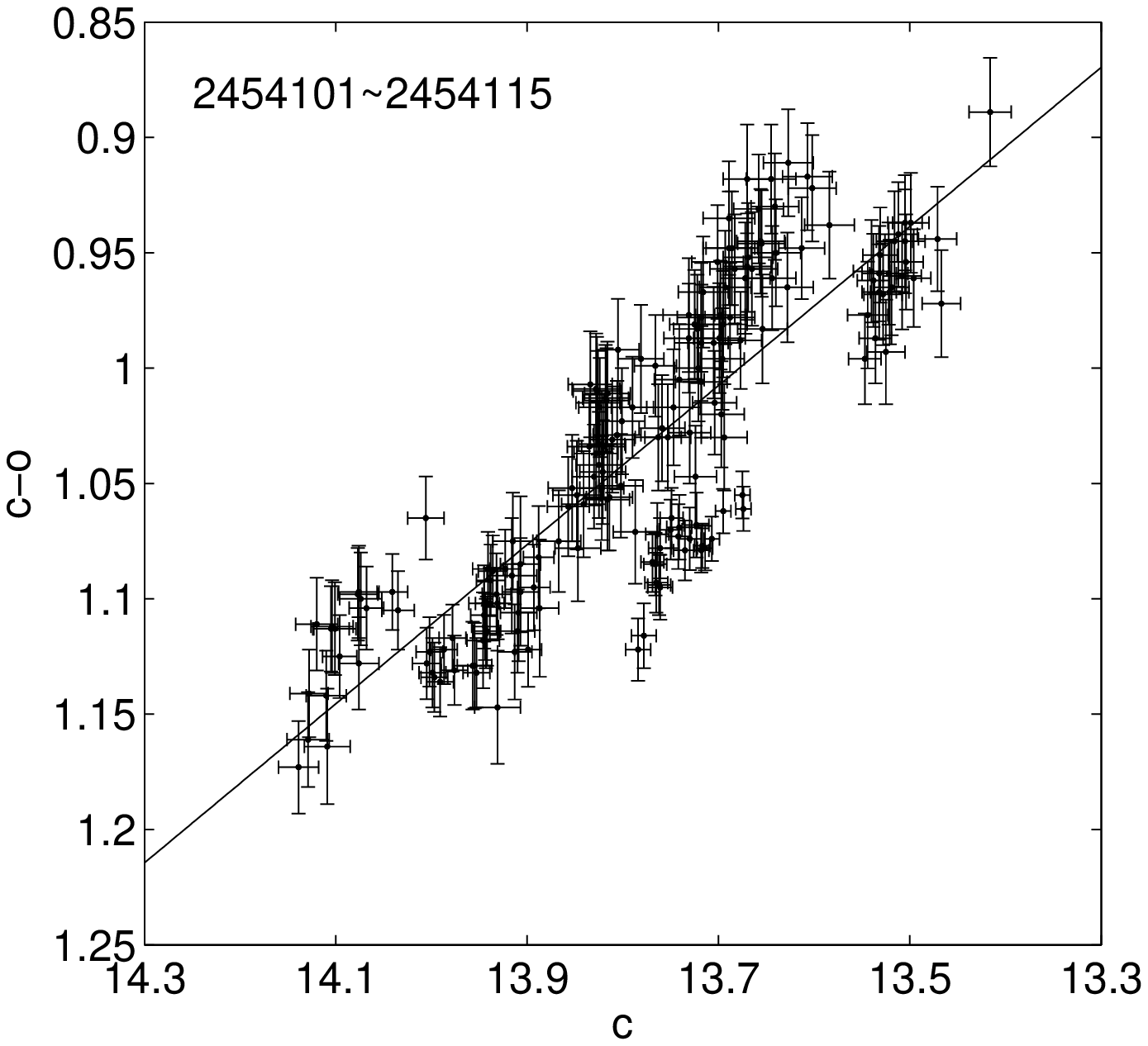}
\includegraphics[angle=0,width=0.48\textwidth]{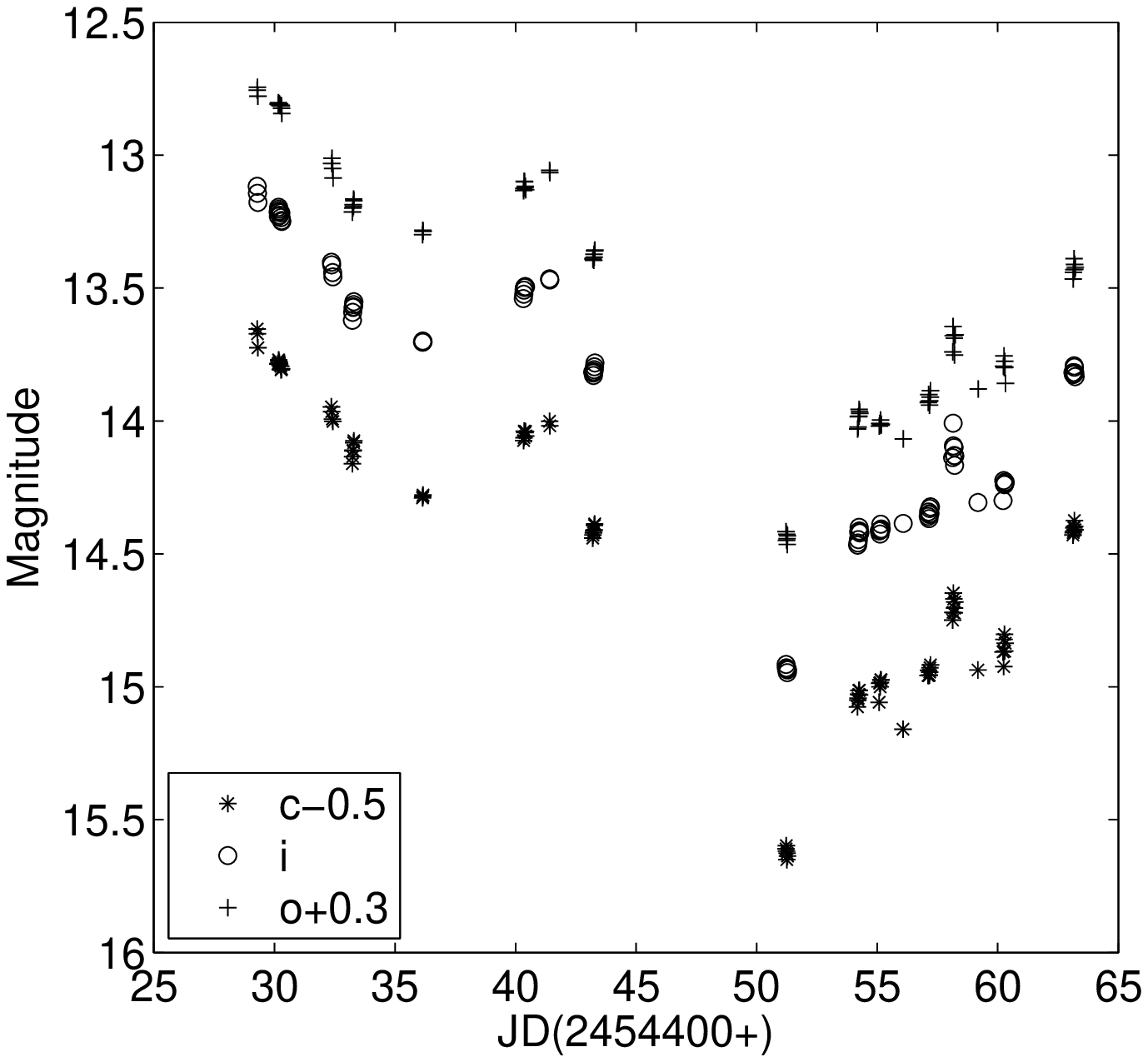}
\includegraphics[angle=0,width=0.48\textwidth]{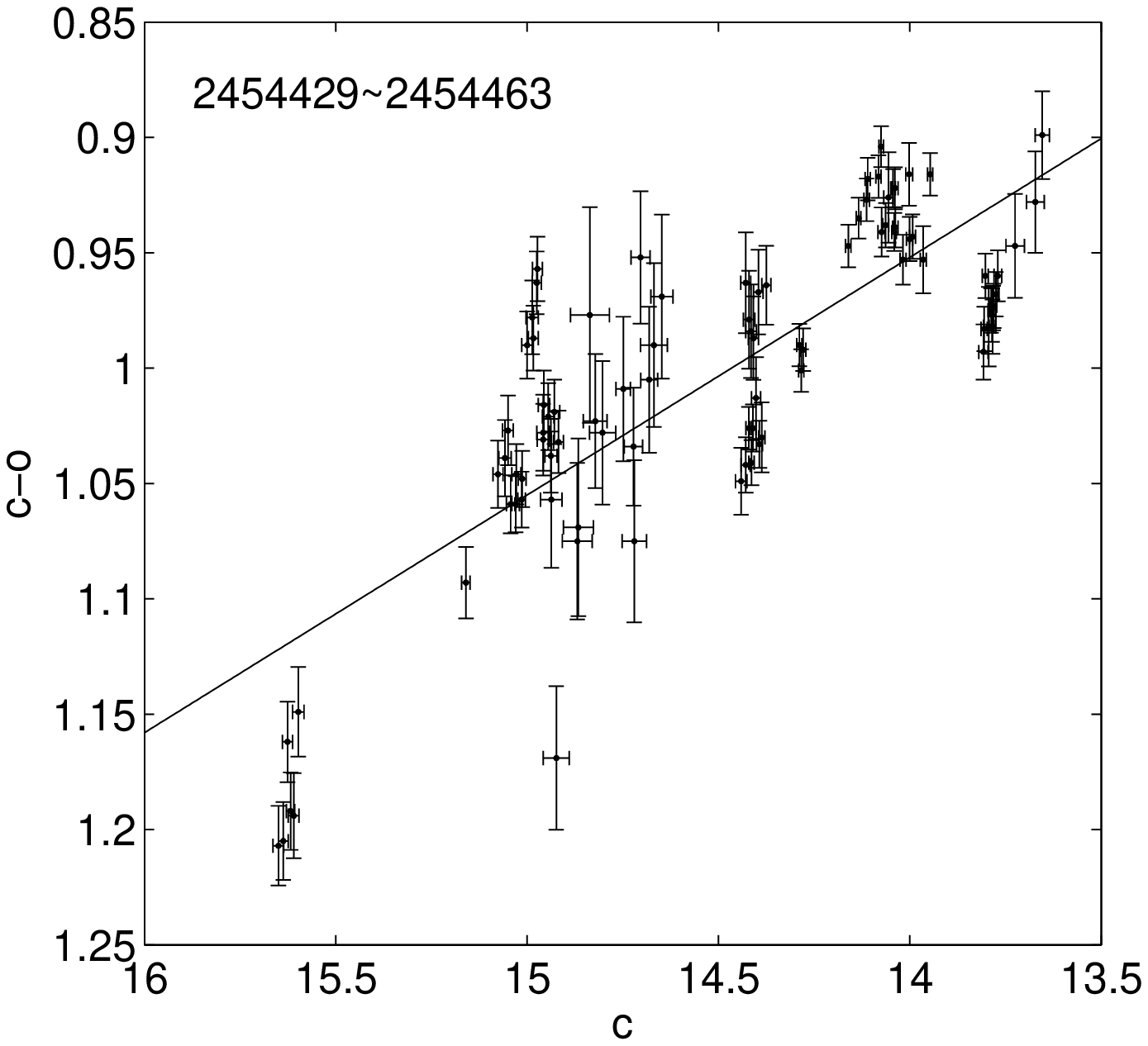}
\includegraphics[angle=0,width=0.48\textwidth]{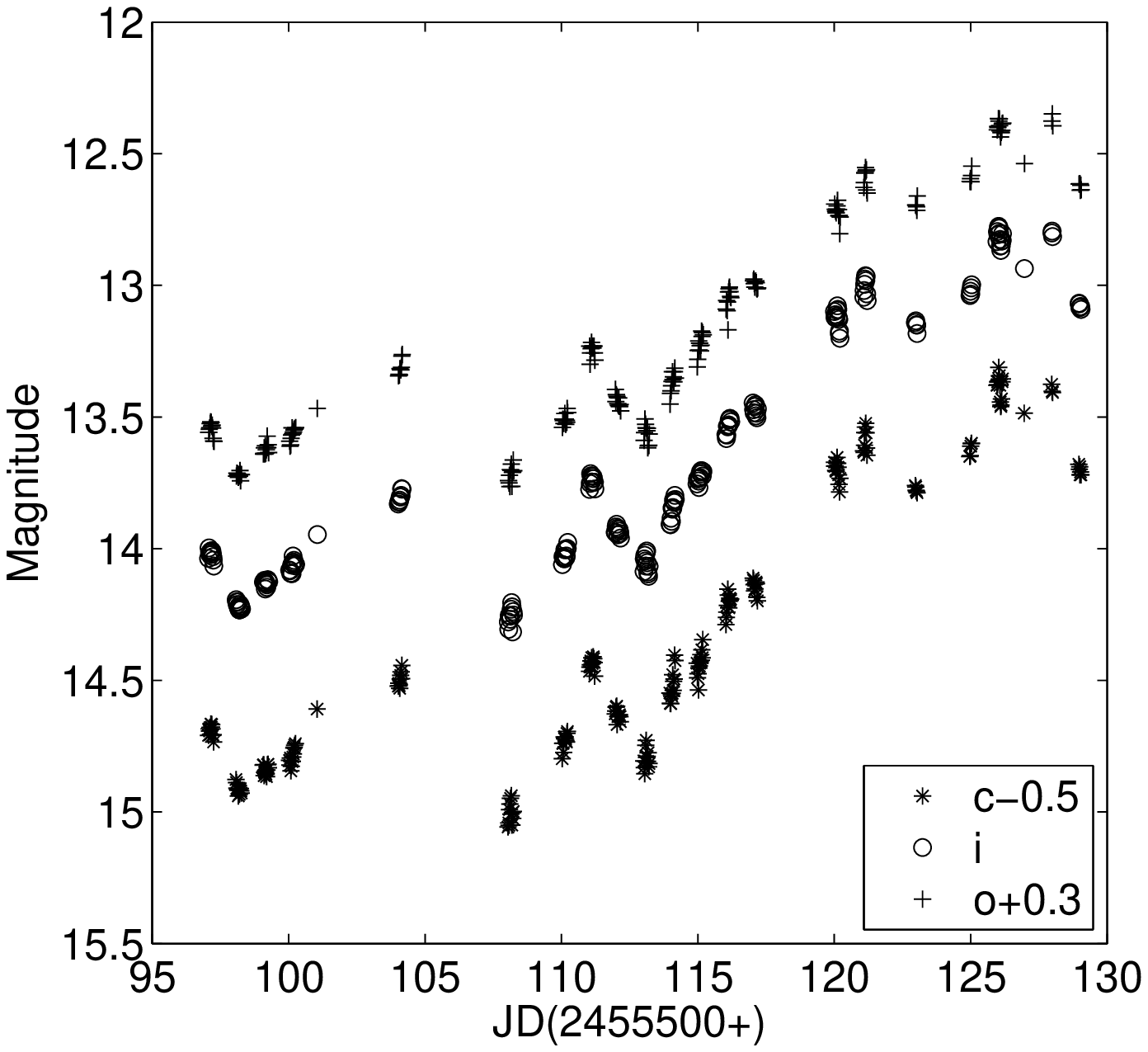}
\includegraphics[angle=0,width=0.48\textwidth]{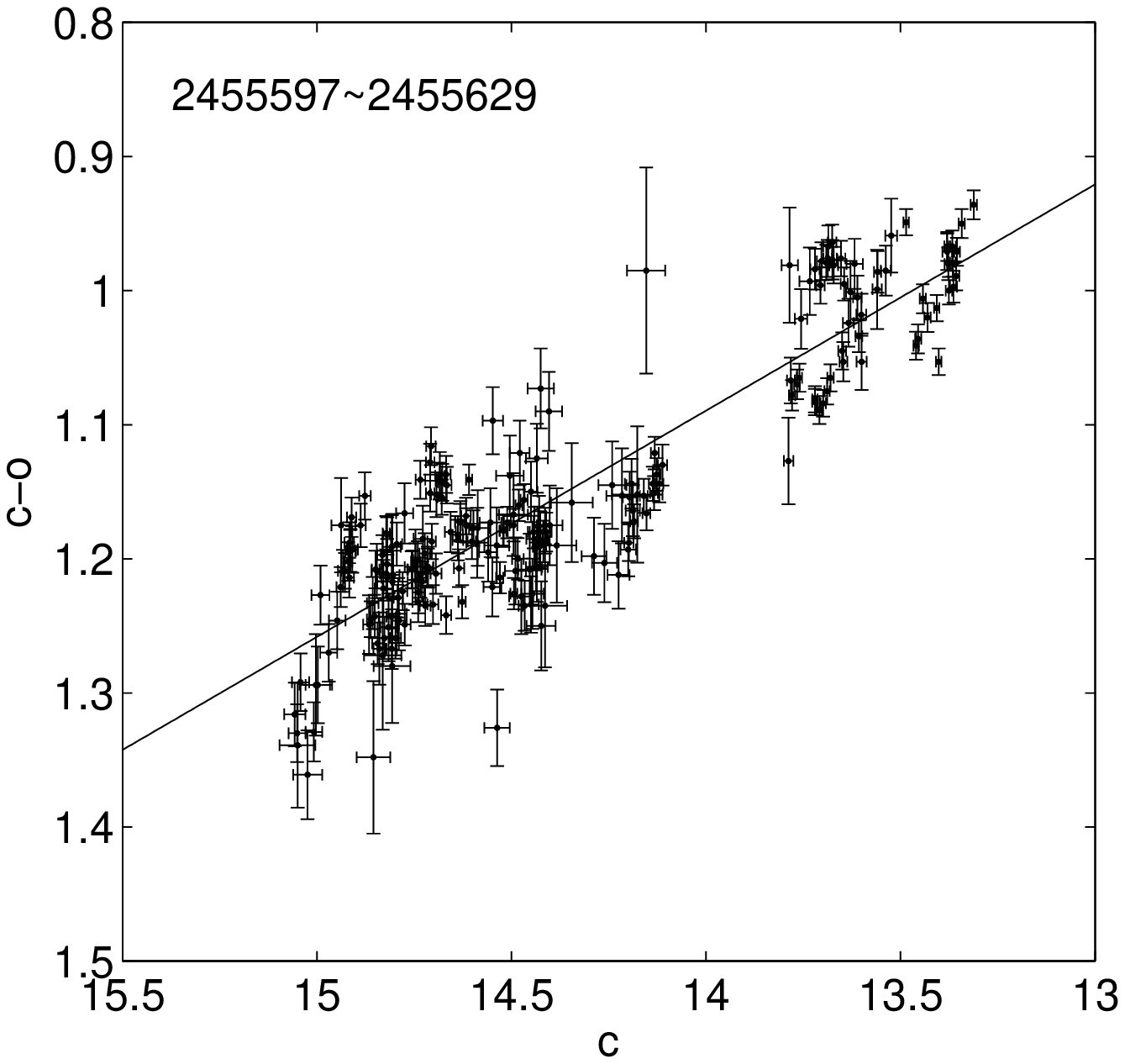}
\caption{Light curves (left) and Color-magnitude diagrams (right)
of intermediate timescales. The line in the
right figures is the linear fit to the data. \label{fig6}}
\end{center}
\end{figure}

\begin{figure}
\begin{center}
\includegraphics[angle=0,width=0.48\textwidth]{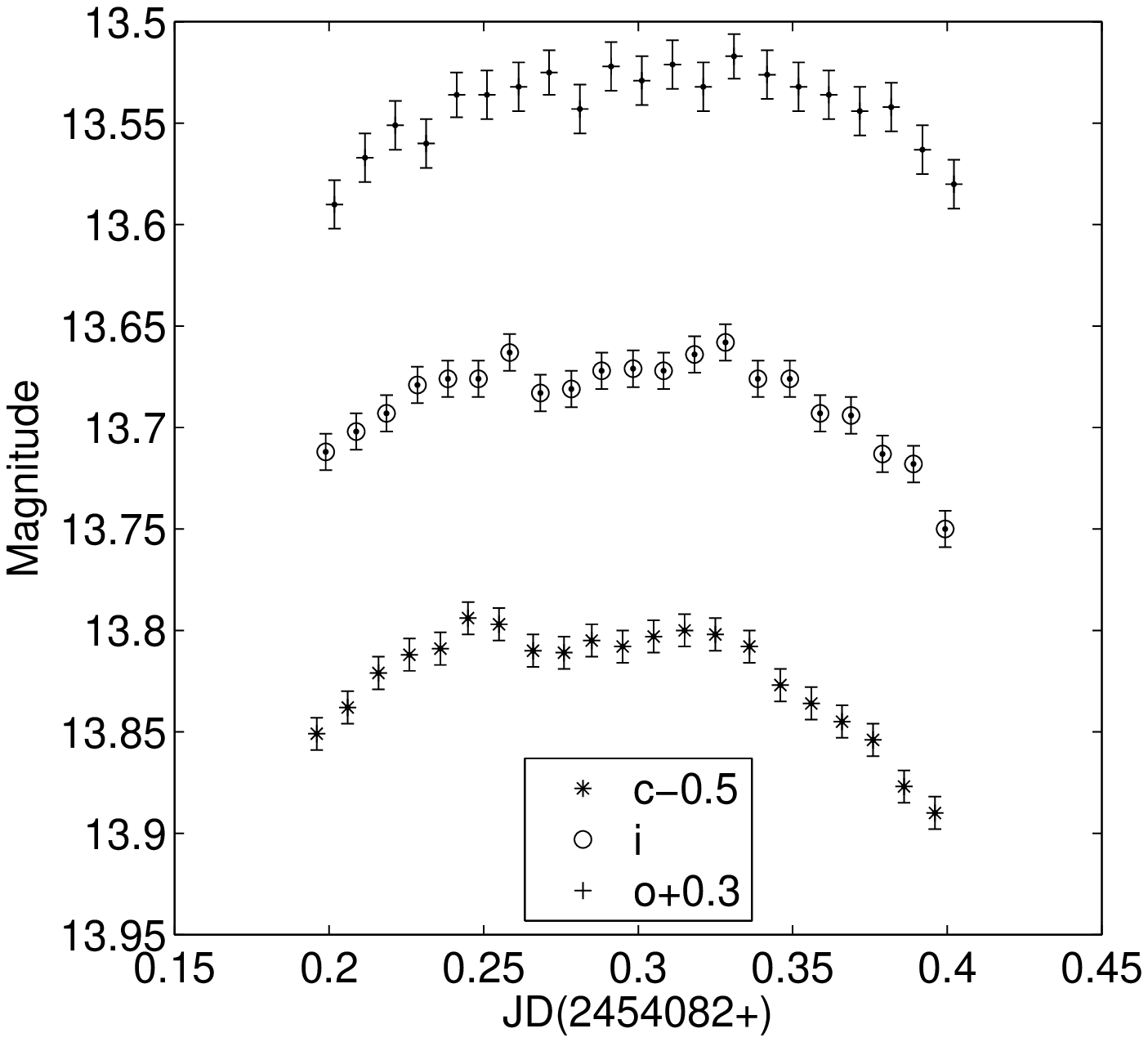}
\includegraphics[angle=0,width=0.48\textwidth]{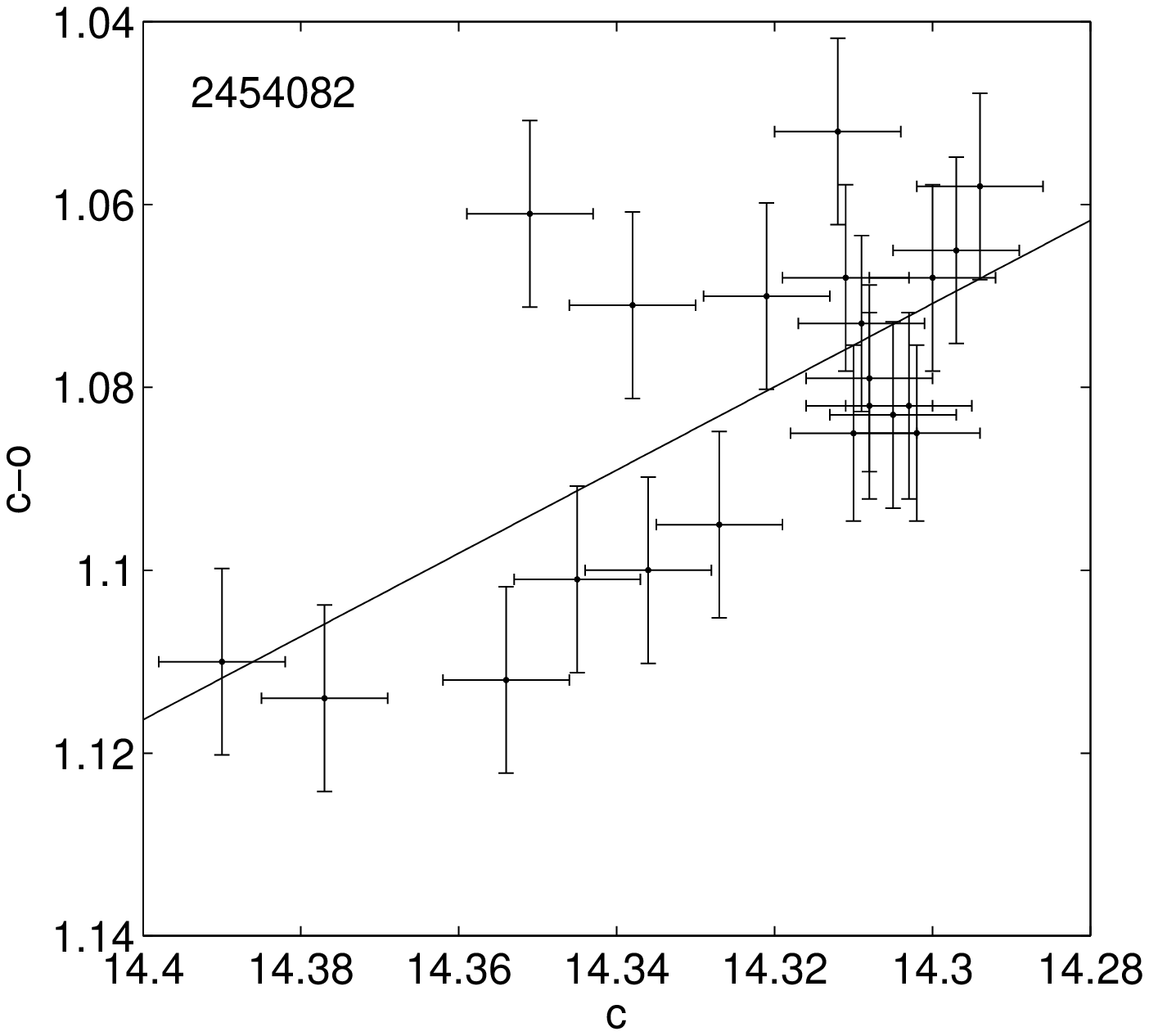}
\includegraphics[angle=0,width=0.48\textwidth]{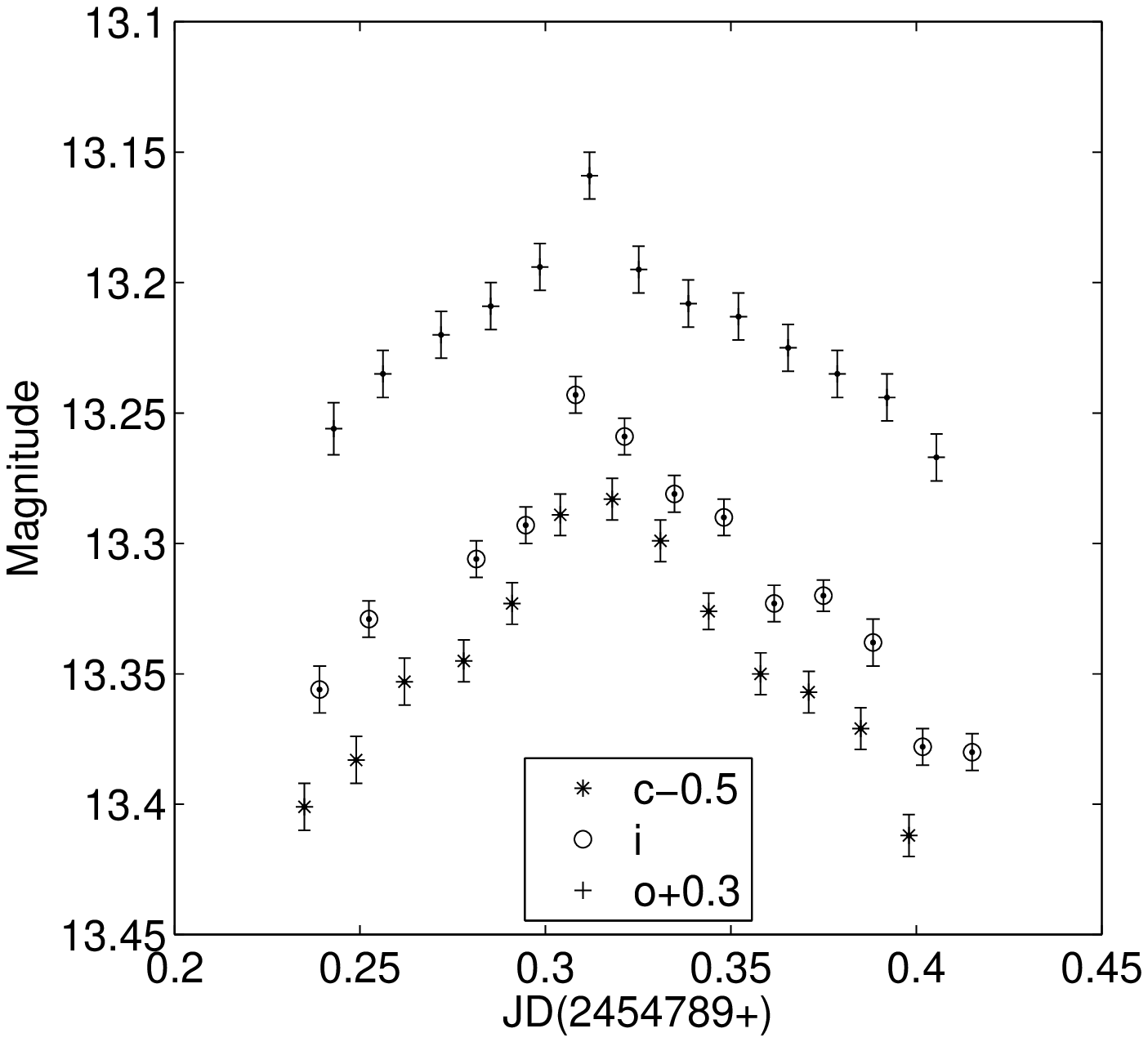}
\includegraphics[angle=0,width=0.48\textwidth]{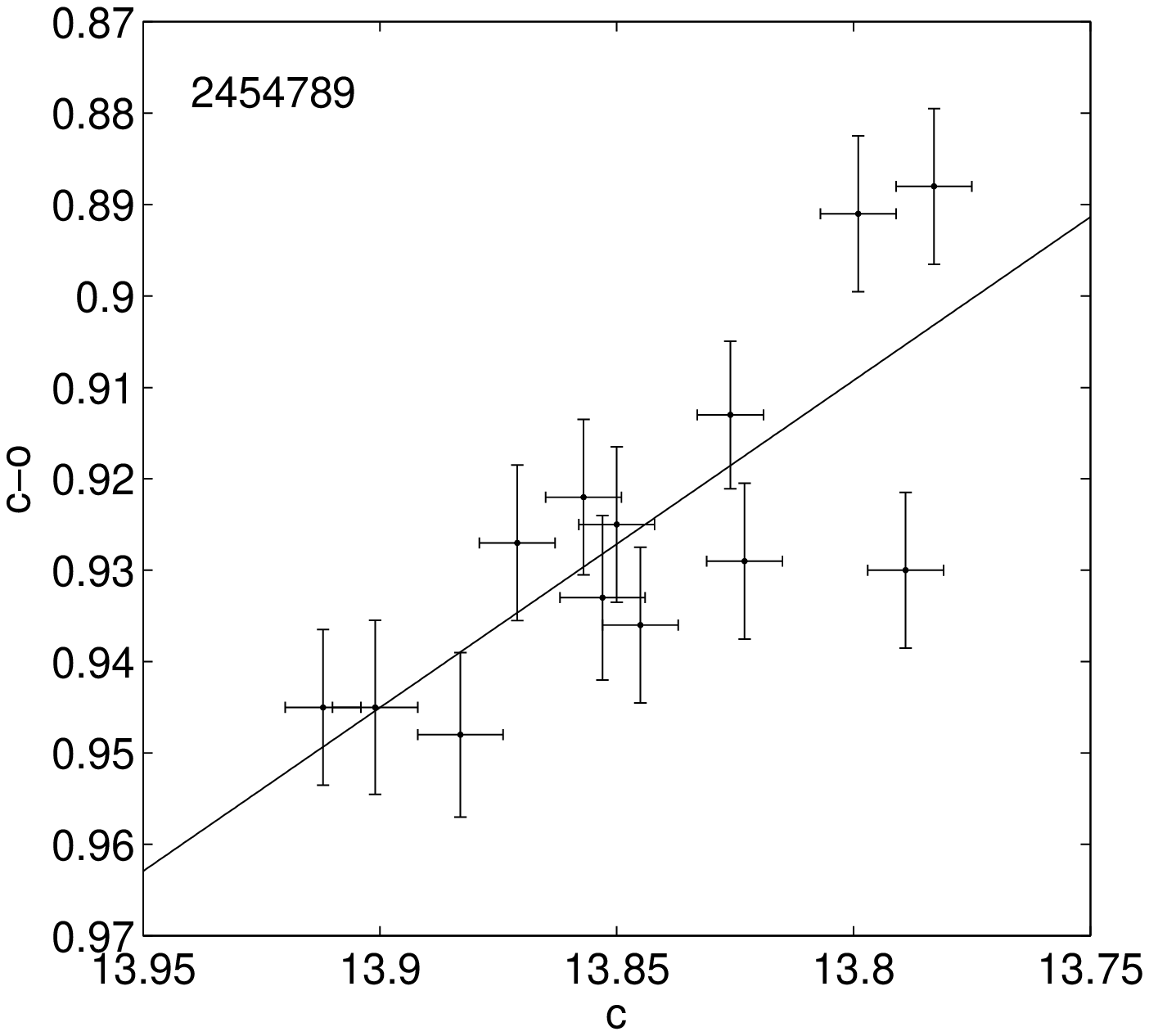}
\includegraphics[angle=0,width=0.48\textwidth]{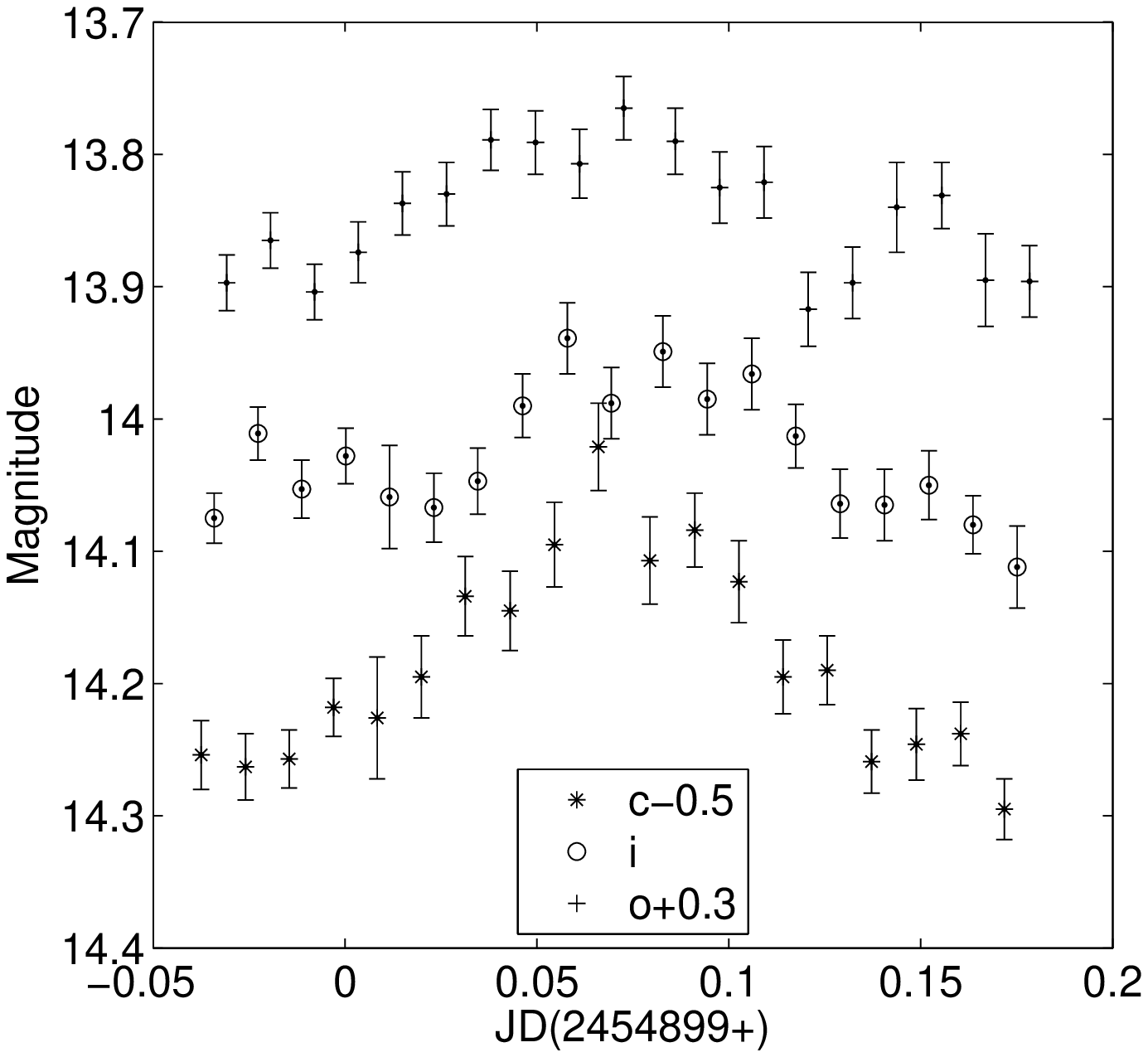}
\includegraphics[angle=0,width=0.48\textwidth]{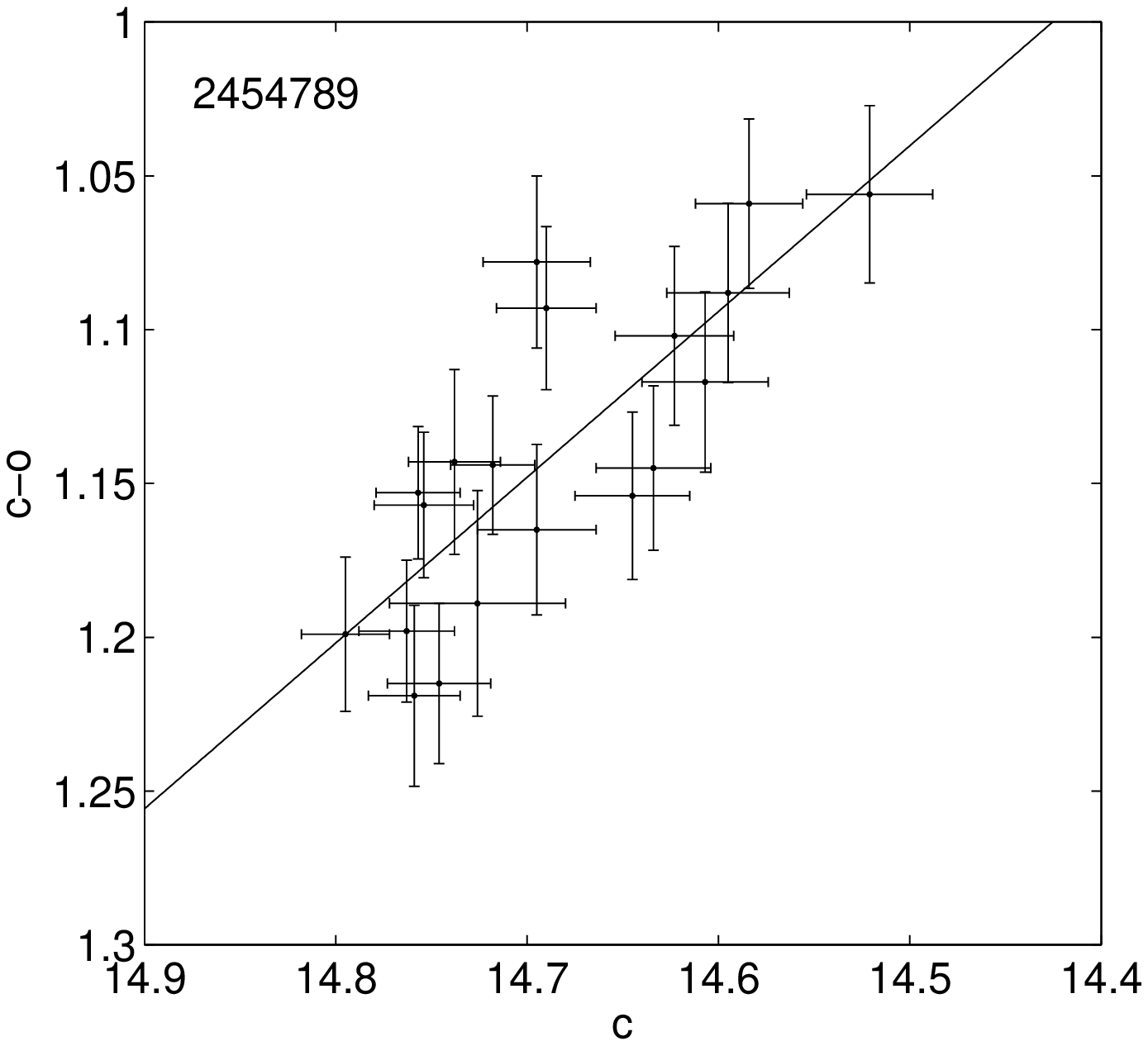}
\caption{Light curves (left) and Color-magnitude diagrams (right)
of IDV data. The line in the right figures
is the linear fit to the data. \label{fig7}}
\end{center}
\end{figure}

\begin{figure}
\begin{center}
\includegraphics[angle=0,width=0.48\textwidth]{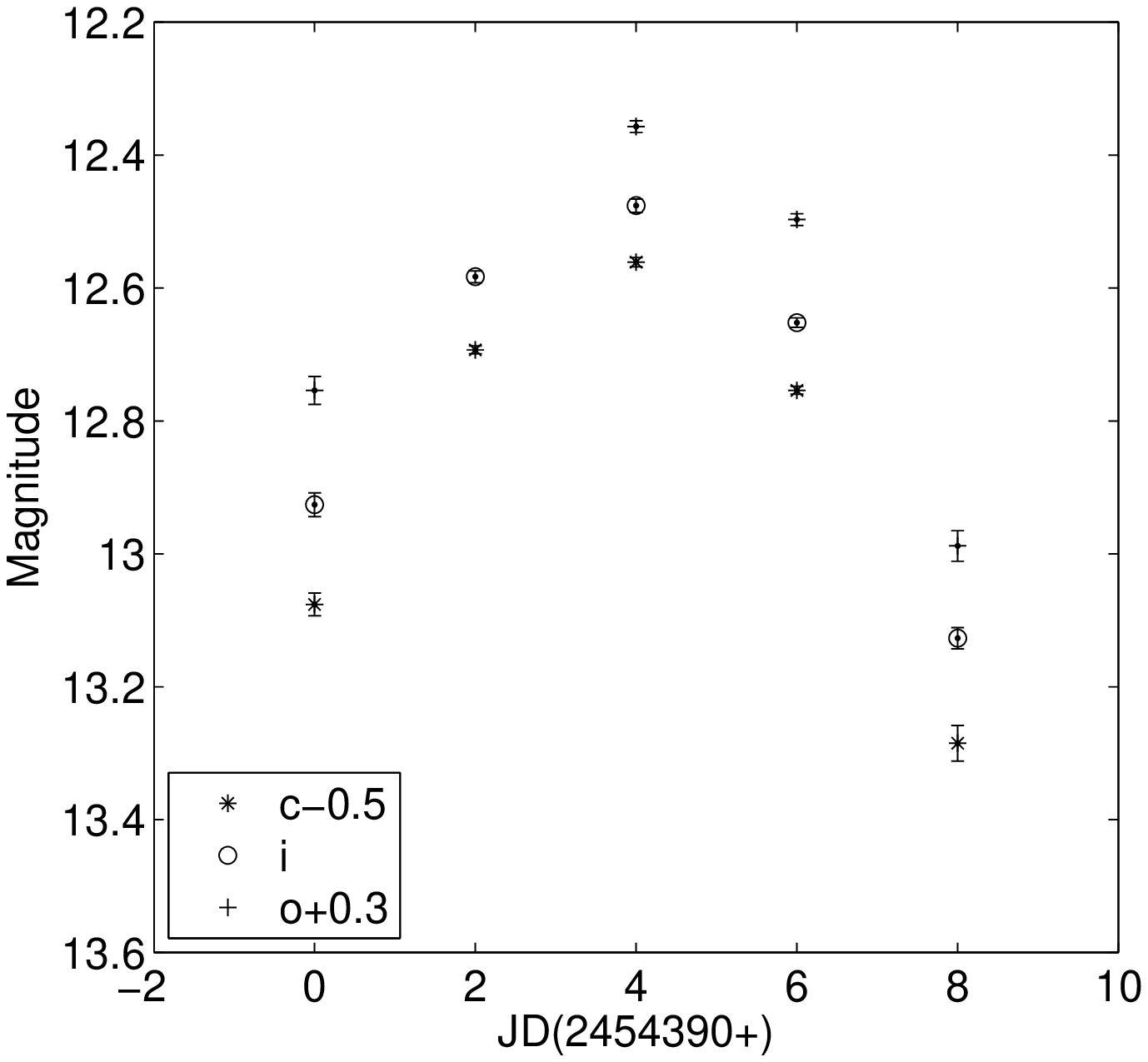}
\includegraphics[angle=0,width=0.48\textwidth]{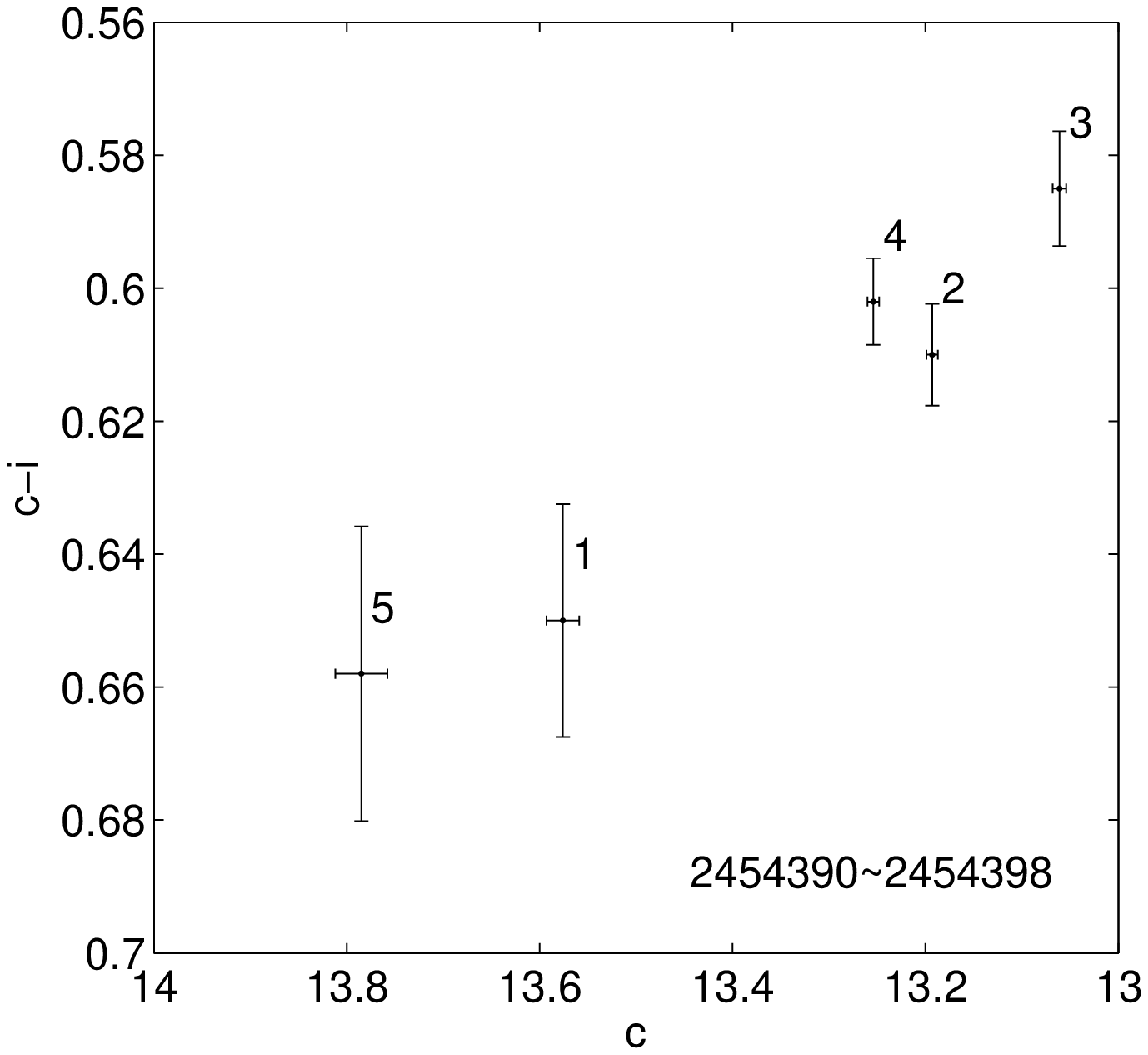}
\includegraphics[angle=0,width=0.48\textwidth]{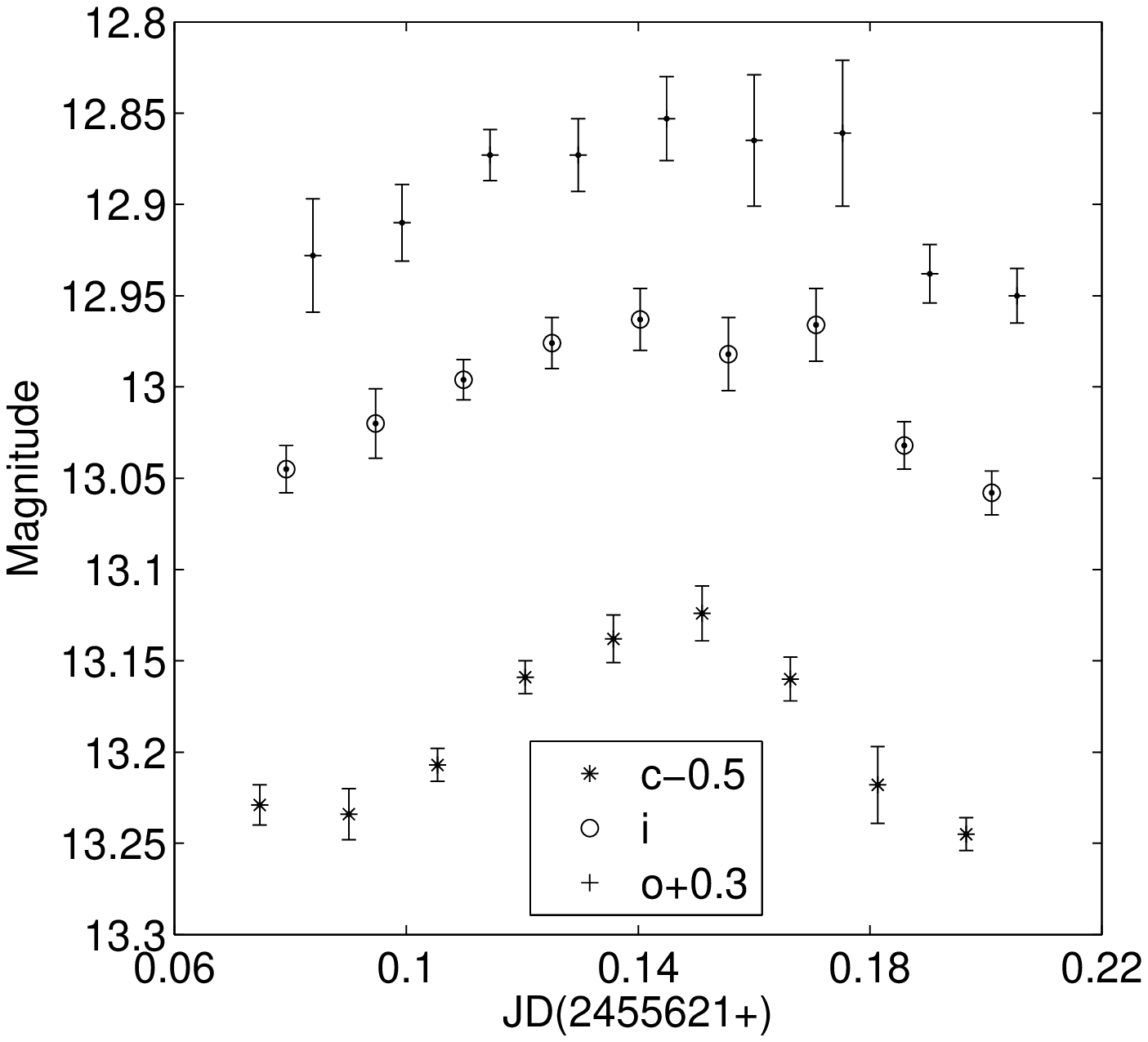}
\includegraphics[angle=0,width=0.48\textwidth]{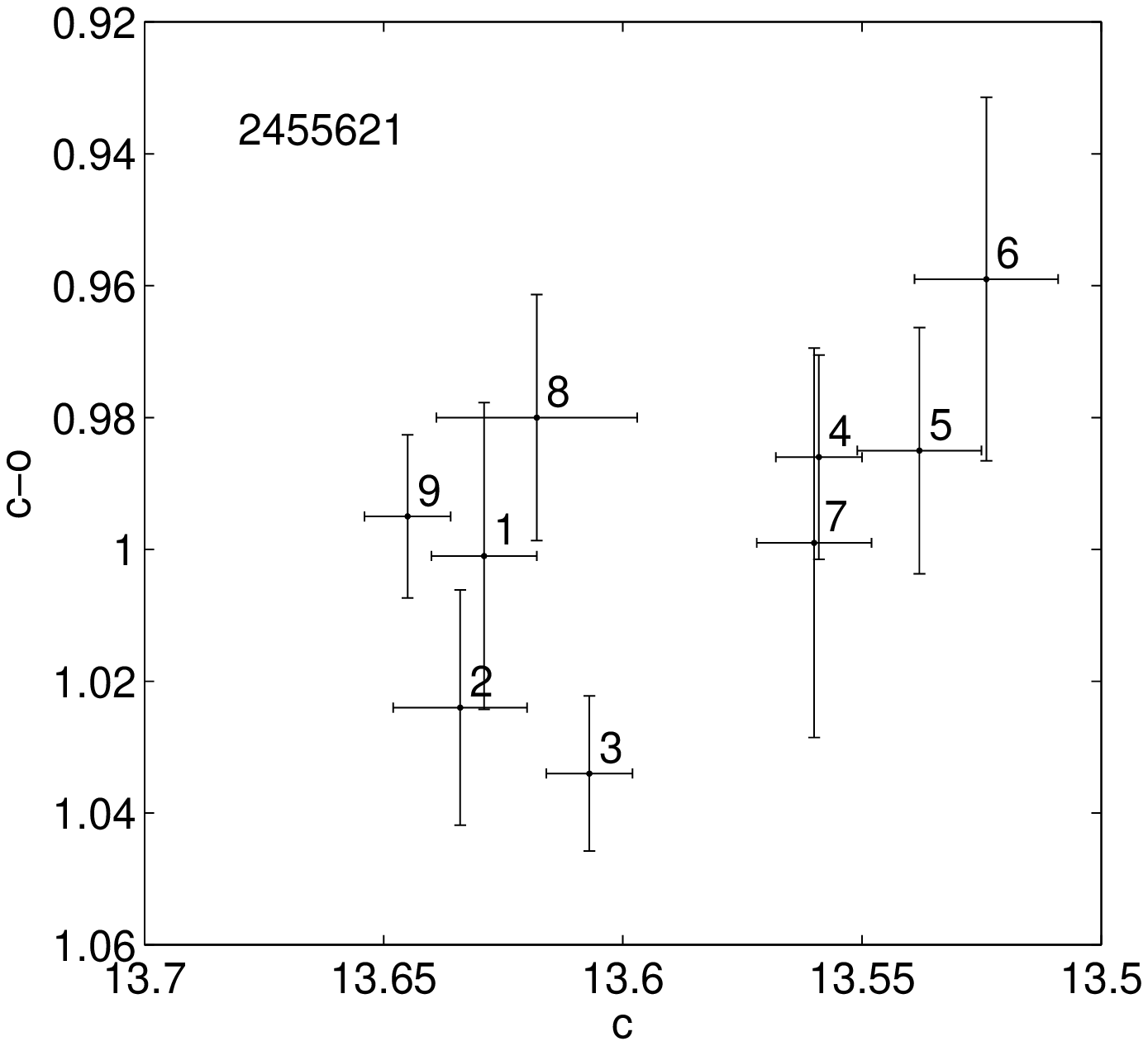}
\caption{Light curves (left) and color-magnitude diagrams (right)
for the internight (upper) and intranight (lower) timescales.
The numbers in the right panels denote the time sequence.\label{fig8}}
\end{center}
\end{figure}

\clearpage

\begin{deluxetable}{cccccc}
\tablecaption{Magnitudes of 4 Comparison Stars.\label{tbl-1}} \tablewidth{0pt}
\tablehead{\colhead{Star} & \colhead{$c$} & \colhead{$e$} & \colhead{$i$} & \colhead{$m$} & \colhead{$o$}} \startdata
 3 & 13.033 & 12.706 & 12.271 & 12.190 & 12.178 \\
 4 & 13.574 & 13.412 & 13.098 & 13.050 & 13.060 \\
 5 & 14.126 & 13.794 & 13.363 & 13.264 & 13.223 \\
 6 & 14.195 & 13.890 & 13.442 & 13.370 & 13.329 \\
\enddata
\end{deluxetable}

\begin{deluxetable}{ccccccr}
\tablecaption{Data of $c$ band\label{tbl-2}} \tablewidth{0pt}
\tablehead{\colhead{Date(UT)} & \colhead{Time} & \colhead{Julian
Date} & \colhead{Exp} & \colhead{$c$} & \colhead{$c_{err}$} &
\colhead{dfmag}} \startdata
2006 12 06 &18:43:43.0 &2454076.28036 &300 &14.649 &0.016 &0.002 \\
2006 12 06 &18:58:33.0 &2454076.29066 &300 &14.636 &0.020 &$-$0.018 \\
2006 12 06 &19:12:43.0 &2454076.30050 &300 &14.639 &0.014 &0.011 \\
2006 12 06 &19:27:06.0 &2454076.31049 &300 &14.660 &0.014 &0.004 \\
2006 12 06 &19:41:42.0 &2454076.32063 &300 &14.612 &0.014 &$-$0.003 \\
\enddata
\tablecomments{Table \ref{tbl-2} is published in its entirety in the
electronic edition of the {\sl The Astrophysical Journal Supplement}. A portion is
shown here for guidance regarding its form and content.}
\end{deluxetable}

\begin{deluxetable}{ccccccr}
\tablecaption{Data of $e$ band\label{tbl-3}} \tablewidth{0pt}
\tablehead{\colhead{Date(UT)} & \colhead{Time} & \colhead{Julian
Date} & \colhead{Exp} & \colhead{$e$} & \colhead{$e_{err}$} &
\colhead{dfmag}} \startdata
2005 01 28 &15:57:45.0 &2453399.16510 &300 &14.216 &0.027 &$-$0.009 \\
2005 01 28 &16:07:44.0 &2453399.17204 &300 &14.212 &0.026 &$-$0.028 \\
2005 01 28 &16:17:43.0 &2453399.17897 &300 &14.186 &0.023 &0.015 \\
2005 01 28 &16:27:57.0 &2453399.18608 &300 &14.239 &0.029 &$-$0.015 \\
2005 01 28 &16:37:53.0 &2453399.19297 &300 &14.242 &0.026 &0.011 \\
\enddata
\tablecomments{Table \ref{tbl-3} is published in its entirety in the
electronic edition of the {\sl The Astrophysical Journal Supplement}. A portion is
shown here for guidance regarding its form and content.}
\end{deluxetable}

\begin{deluxetable}{ccccccr}
\tablecaption{Data of $i$ band\label{tbl-4}} \tablewidth{0pt}
\tablehead{\colhead{Date(UT)} & \colhead{Time} & \colhead{Julian
Date} & \colhead{Exp} & \colhead{$i$} & \colhead{$i_{err}$} &
\colhead{dfmag}} \startdata
2004 12 20 &17:45:00.0 &2453360.23958 &240 &12.739 &0.011 &$-$0.012 \\
2004 12 20 &17:51:11.0 &2453360.24388 &240 &12.745 &0.010 &0.010 \\
2004 12 20 &17:55:40.0 &2453360.24699 &240 &12.717 &0.010 &$-$0.026 \\
2004 12 20 &18:00:06.0 &2453360.25007 &240 &12.748 &0.011 &0.032 \\
2004 12 20 &18:04:36.0 &2453360.25319 &240 &12.733 &0.014 &$-$0.008 \\
\enddata
\tablecomments{Table \ref{tbl-4} is published in its entirety in the
electronic edition of the {\sl The Astrophysical Journal Supplement}. A portion is
shown here for guidance regarding its form and content.}
\end{deluxetable}

\begin{deluxetable}{ccccccr}
\tablecaption{Data of $m$ band\label{tbl-5}} \tablewidth{0pt}
\tablehead{\colhead{Date(UT)} & \colhead{Time} & \colhead{Julian
Date} & \colhead{Exp} & \colhead{$m$} & \colhead{$m_{err}$} &
\colhead{dfmag}} \startdata
2005 12 21 &18:46:03.0 &2453726.28198 &300 &12.549 &0.007 &$-$0.019 \\
2005 12 21 &19:02:20.0 &2453726.29329 &300 &12.550 &0.007 &$-$0.013 \\
2005 12 21 &19:18:14.0 &2453726.30433 &300 &12.551 &0.007 &0.012 \\
2005 12 21 &19:34:46.0 &2453726.31581 &300 &12.545 &0.008 &0.007 \\
2005 12 21 &19:50:41.0 &2453726.32686 &300 &12.537 &0.008 &$-$0.002 \\
\enddata
\tablecomments{Table \ref{tbl-5} is published in its entirety in the
electronic edition of the {\sl The Astrophysical Journal
Supplement}. A portion is shown here for guidance regarding its form
and content.}
\end{deluxetable}

\begin{deluxetable}{ccccccr}
\tablecaption{Data of $o$ band\label{tbl-6}} \tablewidth{0pt}
\tablehead{\colhead{Date(UT)} & \colhead{Time} & \colhead{Julian
Date} & \colhead{Exp} & \colhead{$o$} & \colhead{$o_{err}$} &
\colhead{dfmag}} \startdata
2006 12 06 &18:51:49.0 &2454076.28598 &300 &13.580 &0.018 &0.004 \\
2006 12 06 &19:06:42.0 &2454076.29632 &300 &13.555 &0.017 &0.007 \\
2006 12 06 &19:21:04.0 &2454076.30630 &300 &13.557 &0.016 &0.008 \\
2006 12 06 &19:35:38.0 &2454076.31641 &300 &13.562 &0.016 &$-$0.005 \\
2006 12 06 &19:50:16.0 &2454076.32657 &300 &13.532 &0.016 &0.004 \\
\enddata
\tablecomments{Table \ref{tbl-6} is published in its entirety in the
electronic edition of the {\sl The Astrophysical Journal Supplement}. A portion is
shown here for guidance regarding its form and content.}
\end{deluxetable}

\end{document}